\documentclass[journal,twoside]{IEEEtran}
\usepackage{url}
\usepackage[sort,nocompress]{cite}
\usepackage{amsmath,amssymb,amsfonts}
\usepackage{xcolor}
\usepackage[ruled,linesnumbered,vlined]{algorithm2e}
\usepackage{array}
\usepackage{textcomp}
\usepackage{stfloats}
\usepackage{verbatim}
\usepackage{graphicx}
\usepackage{bbm}
\usepackage{multirow}
\usepackage{leftindex}
\usepackage{latexsym}
\usepackage{mathrsfs}
\usepackage{setspace}
\usepackage{booktabs}
\usepackage{color}
\usepackage{comment}
\usepackage[font=footnotesize,labelsep=period]{caption}
\usepackage{makecell}
\usepackage{threeparttable}
\usepackage{colortbl}
\usepackage{float}
\usepackage{lipsum}
\usepackage{bm}
\usepackage{balance}
\usepackage{cases}

\makeatletter
\renewcommand{\maketag@@@}[1]{\hbox{\m@th\normalsize\normalfont#1}}%
\makeatother
\usepackage[subrefformat=parens,justification=centering]{subcaption}
\newcolumntype{P}[1]{>{\centering\arraybackslash}m{#1}}
\usepackage[colorlinks=true,allcolors=blue,urlcolor=black]{hyperref}

\let\oldeqref\eqref
\renewcommand{\eqref}[1]{\textcolor{blue}{\oldeqref{#1}}}

\usepackage{orcidlink}
\newcommand{\IEEEorcidlink}[1]{\raisebox{1mm}{\kern0.11mm\large\orcidlink{#1}}}

\newcommand{\T}{\mathsf{T}}
\newcommand{\m}{\,\mathrm{m}}

\newcommand{\softmax}{\mathop{\mathrm{softmax}}}

\begin{document}

\title{Fusing Bluetooth With Pedestrian Dead Reckoning: A Floor Plan-Assisted Positioning Approach
}%

\author{Wenxuan~Pan\IEEEorcidlink{0009-0000-7922-1583}, Yang~Yang,~\IEEEmembership{Senior~Member,~IEEE}, Mingzhe~Chen,~\IEEEmembership{Senior~Member,~IEEE}, Dong~Wei, Caili~Guo,~\IEEEmembership{Senior~Member,~IEEE}, and~Shiwen~Mao,~\IEEEmembership{Fellow,~IEEE}%
    \thanks{Received 15 April, 2025. This work was supported in part by National Natural Science Foundation of China under Grant 62371065 and Grant 61871047, and in part by BUPT Innovation and Entrepreneurship Support Program under Grant 2025-YC-S007. An earlier version\cite{MyConf} of this paper was presented at the 2025 IEEE International Conference on Communications Workshops (ICC Workshops) [DOI: 10.1109/ICCWorkshops67674.2025.11162482]. \textit{(Corresponding author: Yang Yang.)}}%
    \thanks{Wenxuan Pan and Yang Yang are with Beijing Key Laboratory of Network System Architecture and Convergence, School of Information and Communication Engineering, Beijing University of Posts and Telecommunications, Beijing 100876, China (e-mail: \mbox{pwx@bupt.edu.cn}; \mbox{yangyang01@bupt.edu.cn}).}%
    \thanks{Mingzhe Chen is with the Department of Electrical and Computer Engineering and the Institute for Data Science and Computing, University of Miami, Coral Gables, FL 33146, USA (e-mail: \mbox{mingzhe.chen@miami.edu}).}%
    \thanks{Dong Wei is with the Institute of Information Engineering, Chinese Academy of Sciences, Beijing 100093, China (e-mail: \mbox{weidong@iie.ac.cn}).}%
    \thanks{Caili Guo is with Beijing Laboratory of Advanced Information Networks, School of Information and Communication Engineering, Beijing University of Posts and Telecommunications, Beijing 100876, China (e-mail: \mbox{guocaili@bupt.edu.cn}).}%
    \thanks{Shiwen Mao is with the Wireless Engineering Research and Education Center, Auburn University, Auburn, AL 36849, USA (e-mail: \mbox{smao@ieee.org}).}%
    \vspace{-0.5cm}%
}

\markboth{IEEE Journal of \LaTeX\ Class Files,~Vol.~14, No.~8, November~2025}%
{Pan \MakeLowercase{et al.}: A Sample Article Using IEEEtran.cls for IEEE Journals}

\IEEEpubid{0000-0000~\copyright~2025 IEEE.}

\maketitle

\begin{abstract}
Floor plans can provide valuable prior information that helps enhance the accuracy of indoor positioning systems. However, existing research typically faces challenges in efficiently leveraging floor plan information and applying it to complex indoor layouts. To fully exploit information from floor plans for positioning, we propose a floor plan-assisted fusion positioning algorithm (FP-BP) using Bluetooth low energy (BLE) and pedestrian dead reckoning (PDR). In the considered system, a user holding a smartphone walks through a positioning area with BLE beacons installed on the ceiling, and can locate himself in real time. In particular, FP-BP consists of two phases. In the offline phase, FP-BP programmatically extracts map features from a stylized floor plan based on their binary masks, and constructs a mapping function to identify the corresponding map feature of any given position on the map. In the online phase, FP-BP continuously computes BLE positions and PDR results from BLE signals and smartphone sensors, where a novel grid-based maximum likelihood estimation (GML) algorithm is introduced to enhance BLE positioning. Then, a particle filter is used to fuse them and obtain an initial estimate. Finally, FP-BP performs post-position correction to obtain the final position based on its specific map feature. {Experimental results show that FP-BP can achieve a real-time mean positioning accuracy of 1.14 m, representing an improvement of over 29\% compared to existing floor plan-fused baseline algorithms.}
\end{abstract}


\section{Introduction}
\IEEEPARstart{W}{ith} the wide adoption of emerging applications such as smart cities, Industry 4.0, and extended reality (XR), the need for accurate indoor positioning has become more critical than ever. Although the global navigation satellite systems (GNSSs) have demonstrated high accuracy and wide coverage in outdoor scenarios, they still face challenges indoors due to the weak satellite signals and environmental obstructions\cite{Yang2024,Koelemeij2022}. To address these challenges, various indoor positioning technologies, such as WiFi\cite{Caso2020}, Bluetooth low energy (BLE)\cite{Lee2019}, radio frequency identification (RFID)\cite{Zhang2019RFID}, visible light positioning (VLP)\cite{pan2025VLP}, and ultra-wideband (UWB)\cite{Zhang2020}, have been proposed. {Among these, both WiFi and BLE share advantages such as low cost and wide availability in commercial devices. Nevertheless, BLE tends to offer additional practical advantages, owing to its lower power consumption\cite{Yang2024,Jeon2018}, higher deployment flexibility\cite{Jeon2018,Zhuang2022} and generally improved positioning accuracy\cite{Farahsari2022}.}

BLE positioning typically uses the received signal strength indicator (RSSI) as the primary measurement, deriving common algorithms such as trilateration\cite{Spachos2020} and fingerprinting\cite{Ng2023,Faragher2015}. The ease of obtaining BLE RSSIs has facilitated the widespread adoption of RSSI-based algorithms in BLE positioning. However, due to the complexity of indoor environments, the BLE RSSI is affected by environmental variations and interference, resulting in significant fluctuations, limiting RSSI-based BLE positioning to meter-level accuracy\cite{Zhuang2022,Farahsari2022}. Since Bluetooth 5.1\cite{BLECore51}, the Bluetooth Special Interest Group (SIG) has introduced new positioning methods based on the angle of arrival (AoA) or angle of departure (AoD). Although AoA/AoD-based algorithms have demonstrated higher accuracy than those based on RSSI in general\cite{Yang2024}, the additional multi-antenna hardware requirements also lead to increased positioning costs\cite{Huang2021}. More importantly, common devices, such as smartphones, are equipped with only one single Bluetooth antenna, which makes AoA/AoD-based algorithms inapplicable. Therefore, RSSI-based algorithms remain the most widely used in BLE positioning\cite{Sun2021A}.

\IEEEpubidadjcol

To improve the positioning accuracy of BLE, some fusion-based positioning technologies have been proposed. Among these, pedestrian dead reckoning (PDR) has emerged as a popular method\cite{Hou2021}. Note that PDR typically comprises an inertial measurement unit (IMU) consisting of an accelerometer and a gyroscope, a magnetometer, and other sensors, and thus it can provide abundant extra information, such as movement speed and direction, over short periods. However, PDR requires an accurate initial position and suffers from significant accuracy degradation over time due to cumulative errors of PDR\cite{Hou2021,Zhu2024ASurvey}. By integrating other positioning methods such as BLE, these errors can be effectively mitigated\cite{Zhuang2022}. In particular, the absolute position references provided by BLE can help correct the PDR cumulative errors, and thus enhance the overall reliability of the fusion system. 

Due to the lack of indoor layout information, traditional algorithms inevitably yield results that appear in undesirable areas\cite{Du2019}, such as inside obstacles or cross walls, which compromises their system stability. Under this background, the integration of floor plans as extra prior information, has shown potential. The floor plan can provide serious restrictions of the indoor environment\cite{Du2019,Xu2022}, making floor plan-integration a promising direction for advancing fusion-based positioning algorithms.

\subsection{Related Works}

Current studies on fusing BLE/WiFi with PDR can be categorized into two types: (i) Calculate the BLE/WiFi positioning results first, typically through methods like trilateration\cite{Dinh2021,Poulose2019} or fingerprinting\cite{Du2019,Sun2020}, and then fuse these results with PDR; (ii) Fuse the received BLE/WiFi data, such as BLE RSSI or WiFi round-trip time (RTT), with IMU data directly\cite{Liu2024}. For instance, Kong et al.\cite{Kong2023} directly used an adaptive feedback extended Kalman filter (EKF) to fuse beacon positions, received RSSIs, and the result of PDR, to enhance the accuracy of the estimated position. In terms of filtering mechanisms, the above algorithms commonly employ the Kalman filter (KF)\cite{Dinh2021,Poulose2019} or its extended variations, such as the EKF\cite{Sun2020,Kong2023}, as well as the particle filter (PF)\cite{Du2019}. However, none of the above traditional algorithms consider the integration of floor plans, which contain useful information such as environmental restrictions and pedestrian accessibility for positioning.

To further take the environmental factors into consideration, integrating floor plans in positioning systems has become a research hotspot\cite{Xu2022,Yu2024,Chai2023,Huang2020}. In general, the floor plan can also be regarded as a type of sensor\cite{Zhuang2022}. In indoor scenarios, floor plans are readily available\cite{Yu2024} and can effectively constrain positioning results to the walkable area, thereby improving system performance\cite{Guo2020,Sui2024}. With respect to the filtering mechanisms, the PF is highly compatible with map information\cite{Xu2022}. As such, many floor plan-integrated algorithms adopt PF for fusion. In\cite{Nurminen2016}, the authors generate particles in PF based on a specific angular probability density function (PDF), effectively avoiding ``wall-crossing” particles. However, this algorithm requires frequent and extensive wall-distance checks, resulting in high computational cost in practical scenarios. The authors in\cite{Huang2020} and\cite{Xia2019} directly assign zero weights to ``wall-crossing” particles in PF. Similarly, in\cite{Sun2025}, the authors use the map to remove particles that fall outside the map boundaries. {However, these algorithms\cite{Huang2020,Xia2019,Sun2025} neglect implicit but crucial information such as the orientations of walls and corners. Moreover, the simple elimination of particles may lead to particle depletion or even positioning failure.} The authors in\cite{Lin2024} determine and identify whether the user has reached a specific corner through gyroscope data to help correct the yaw. {However, this algorithm does not preprocess the floor plan and therefore cannot automatically extract the positions and orientations of corners. As a result, it requires the pre-establishment of a corner database, which inevitably increases the offline workload.} Although interesting, the above algorithms\cite{Huang2020,Xia2019,Sun2025,Lin2024} have not fully utilized the floor plan information, and therefore cannot provide stable long-range positioning or require additional work.

To extract more floor plan information (e.g. wall/corridor directions) for positioning, the authors in\cite{Perttula2014} assume walls are straight lines and determine ``wall-crossing” by checking for intersections between particle movement and walls. Wang et al.\cite{Wang2016An} assume the angles of corners are all right angles, with straight paths between every two adjacent corners. Similarly, Choi et al.\cite{Choi2022} also assume corridors as vertical or horizontal segments, and build a virtual graph to correct the user's position and orientation to the door when detecting a ``door-crossing" event. Du et al.\cite{Du2019} proposed an automatic map preprocessing method and an enhanced PF algorithm that integrates a floor plan based on image processing for line detection. However, the above algorithms\cite{Du2019,Perttula2014,Wang2016An,Choi2022} require the floor plan to exhibit regularly, such as straight corridors with distinct directional changes. This leads to a limitation on map geometry, making them unsuitable for complex floor plans and indoor layouts.

Although the above algorithms\cite{Du2019,Huang2020,Nurminen2016,Xia2019,Sun2025,Lin2024,Perttula2014,Wang2016An} and\cite{Choi2022} have integrated floor plans, they struggle to fully utilize map information or adapt to complex indoor layouts, particularly when extracting map feature boundaries and leveraging their constraints in irregular environments. This may limit the feasibility and accuracy, or increase the deployment cost of the system in real-world scenarios. Thereby, a general and efficient floor plan-assisted fusion positioning algorithm deserves further investigation.

\subsection{Contributions}
{The main contribution of this paper is a novel floor plan-assisted fusion positioning system using BLE and PDR (FP-BP). To the authors’ best knowledge, this is \emph{the first PDR-based framework that integrates floor plans to achieve real-time positioning without limitations on map geometry.}}\footnote{Based on the preliminary version\cite{MyConf} of this work, an enhanced BLE positioning algorithm named GML is proposed in this paper for the first time. Moreover, this paper extends the FP-BP algorithm to multi-floor scenarios to enhance its adaptability and feasibility. Furthermore, based on FP-BP, a prototype is developed for real-time positioning on mobile devices.} Our key contributions are summarized as follows: 
\begin{itemize}
\item {In the online phase, FP-BP first fuses BLE with PDR through the PF to obtain an initial estimation. Then, a post-position correction (PPC) mechanism is proposed to integrate the floor plan deeply into the system, using the map constructed from the floor plan in advance. This mechanism refines the initial estimation based on specific map features to obtain the final result, effectively exploiting the prior knowledge from the floor plan while preventing particle depletion.}
\item {In the offline phase, to programmatically build maps from various floor plans, we propose a feature extraction-based preprocessing method. The only manual step is to first standardize the floor plan by filling obstacle areas with distinguishable colors. Next, a binary mask-based method is applied to extract the inner contours of the floor plan. Finally, a map feature function is defined based on the pre-established planar coordinate systems. This method enables precise pixel-level recognition and extraction of map features, and is therefore widely applicable to various floor plans without restriction.}
\item {To further enhance BLE positioning accuracy, we propose a grid-based maximum likelihood estimation (GML) algorithm. By analyzing the posterior PDF of RSSI-estimated distances, we formulate an objective function based on maximum likelihood estimation (MLE) combined with a novel distance-based penalty term for refinement. Maximizing this function, the algorithm selects the optimal grid point within a virtual grid array as the estimated result, effectively accommodating both dense and sparse beacon deployment scenarios.}
\end{itemize}

Experimental results show that FP-BP algorithm can achieve a mean positioning accuracy of 1.14 m, which is improved by over 29\% compared to existing floor plan-fused baselines. In addition, the proposed GML algorithm also reaches a mean accuracy of 1.74 m when the user keeps moving, which outperforms existing non-fingerprinting baselines by over 34\%.

\subsection{Notations and Paper Organization}

We adopt the following notations: Matrices are denoted by uppercase boldface such as $\bm{R}$; Points and column vectors are denoted by lowercase boldface such as $\bm{v}$; $\bm{v}^\T$ and $\| \bm{v} \|$ represent the transposition and the norm of vector $\bm{v}$, respectively; Sets are denoted by calligraphic characters such as $\mathcal{A}$; $\{ a_i \}_{i=1}^n$ represents a sequence containing $n$ elements and is arranged in order; $\mathcal{N}(\mu,\sigma^2)$ represents a Gaussian distribution with an expectation of $\mu$ and a variance of $\sigma^2$; $p(\bm{x}|\bm{y})$ represents the PDF of a random variable $\bm{x}$ given the condition $\bm{y}$; $\mathbb{E}_{\bm{x}|\bm{y}}[\bm{z}]$ denotes the expectation of a random variable $\bm{z} = \bm{z}(\bm{x})$ under the PDF $p(\bm{x}|\bm{y})$; $\mathbb{T}$ and $\mathbb{F}$ represent Boolean Truth and Falsehood values, respectively.

The rest of the paper is organized as follows. Section~\ref{sec:sys} shows the system model. The proposed GML and FP-BP algorithms are introduced in Sections \ref{sec:ble} and \ref{sec:alg}, respectively. Then, Section~\ref{sec:exp} presents the experimental setup and results of the proposed positioning system. Finally, we conclude this paper in Section~\ref{sec:con}.

\section{System Model and Problem Formulation}
\label{sec:sys}

\subsection{System Model}

This section considers a typical positioning scenario as illustrated in Fig.~\ref{fig:scenario}, where a pedestrian carrying a smartphone passes through an indoor area and receives signals from BLE beacons for positioning. The beacons broadcast BLE signals at a fixed frequency, and the smartphone, i.e., the receiver, continuously scans for these signals. The smartphone also listens to the built-in sensors to obtain the user motion data, which is then fused with the BLE estimated position and the preprocessed floor plan information, allowing real-time active positioning for the user. 

\begin{figure}[!t]
    \centering
    \includegraphics[width=0.485\textwidth]{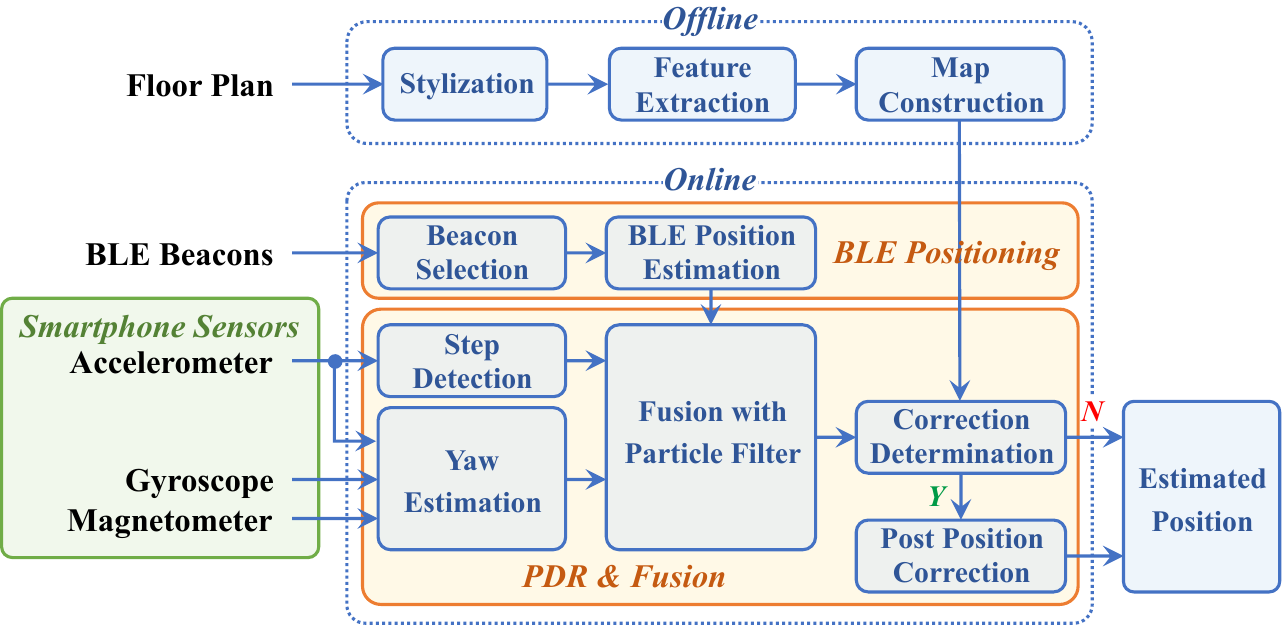}
    \caption{Overall architecture of FP-BP.}
    \label{fig:system}
\end{figure}

The overall architecture of FP-BP is shown in Fig.~\ref{fig:system}. The FP-BP consists of an offline and an online phase. In particular, the online phase of FP-BP further consists of two modules:

\begin{figure}[!t]
    \centering
    \includegraphics[width=0.285\textwidth]{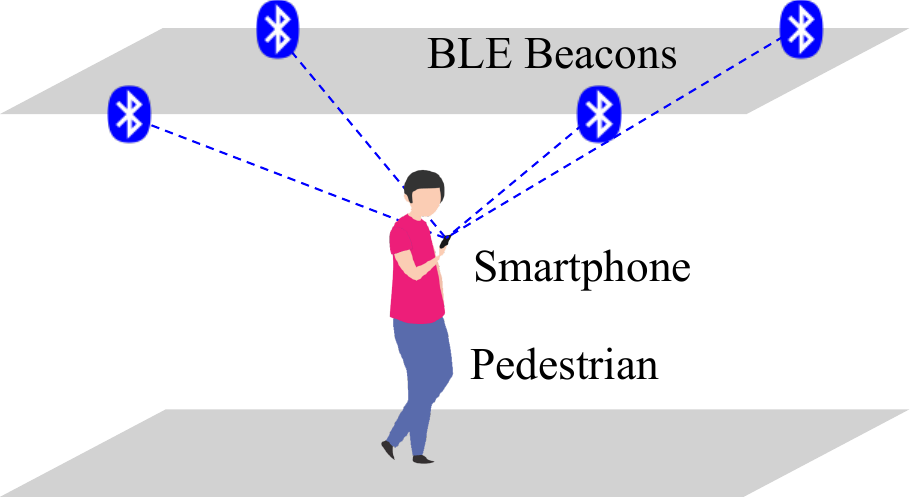}
    \caption{Scenario of the considered system.}
    \label{fig:scenario}
\end{figure}

\subsubsection{BLE Positioning} 
When the smartphone scans for BLE signals, it can obtain information such as the universally unique identifier (UUID) and the instant RSSI of a beacon. The path loss of RSSI can be modeled as\cite{Zhuang2022,Xia2019,Sun2025}: 
\begin{equation}
\label{equ:rssi}
    R = -10n\log d + R_0 + X(\sigma),
\end{equation}
where $R$ is the measured RSSI in dBm; $n$ is the loss factor; $R_0$ is the RSSI at a distance of 1 m, and $X(\sigma) \sim \mathcal{N}(0, \sigma^2)$ is a zero-mean Gaussian noise. Let $\mathcal{B}$ be the set of beacons' positions and $\mathcal{R}$ be the set of received RSSIs. Then, BLE positioning can be expressed as:
\begin{equation}
\label{equ:ble_positioning}
    \bm{x}_B=\bm{f}_{\textit{BLE}}(\mathcal{R}, \mathcal{B}),
\end{equation}
where $\bm{x}_B$ is the estimated position of the user from BLE positioning systems. In particular, we apply the GML algorithm at a fixed time interval in the \textit{BLE Positioning} module to compute and store the current BLE position. This process will be detailed in Section~\ref{sec:ble}.

\subsubsection{PDR\&Fusion}
The smartphone can obtain data from built-in sensors at a fixed frequency for step detection and yaw estimation. Assume the instant sensor data is $\mathcal{D}_s$, and then the position update based on PDR is expressed as:
\begin{equation}
\label{equ:pdr_positioning}
    \bm{x}_P=\bm{x}^{-}+\bm{f}_{\textit{PDR}}(\mathcal{D}_s)\triangleq\bm{x}^{-}+\bm{s},
\end{equation}
where $\bm{x}_P$ is the estimated position of PDR, and $\bm{x}^{-}$ is the positioning result at the previous moment. Upon detecting a step event, i.e., $\|\bm{s}\|>0$, the \textit{PDR\&Fusion} module fuses the step $\bm{s}$ with the stored BLE position using PF algorithm to obtain the initial estimated position. 

Finally, floor plan integration is performed. In the offline phase, floor plan preprocessing has enabled the extraction of the set of points in walkable area, $\mathcal{W}$, from the set of the entire map area, $\mathcal{M}$, in advance. Let the current estimated position be $\widehat{\bm{x}}$. When users walk indoors, they can only move from one walkable position to another, with no obstacles blocking the path. Therefore, we apply the floor plan constraint by ensuring $\bm{x}^{-}$, $\widehat{\bm{x}}$, and all points between these two estimated positions remain within walkable area:
\begin{equation}
\label{equ:pf2}
    (1-\lambda)\bm{x}^-+\lambda \widehat{\bm{x}} \in \mathcal{W},~\forall\lambda\in[0,1].
\end{equation}
In particular, we apply the proposed PPC algorithm to determine and perform necessary correction, and output the final position. This process will be detailed in Section~\ref{sec:alg}.

\subsection{Problem Formulation}
\label{sec:pf}
The target of this paper is to locate the user based on the BLE estimation in \eqref{equ:ble_positioning}, PDR prediction in \eqref{equ:pdr_positioning}, and floor plan constraint \eqref{equ:pf2}. Taking the BLE estimation $\bm{x}_B$ as the current observation, then according to Bayes' theorem, given all observations $\mathcal{Z}$ including $\bm{x}_B$, the posterior PDF of current position $\bm{x}$ is calculated as\cite{Xu2019}:
\begin{equation}
\label{equ:posterior}
    p(\bm{x}|\mathcal{Z}) = \frac{p(\bm{x}_B|\bm{x}) \cdot p(\bm{x}|\mathcal{Z}^-)}{p(\bm{x}_B|\mathcal{Z}^-)},
\end{equation}
where $p(\bm{x}_B|\bm{x})$ is the likelihood; $p(\bm{x}|\mathcal{Z}^-)$ is the prior PDF based on the state transition model; and $p(\bm{x}_B|\mathcal{Z}^-)$ is a normalization constant independent of $\bm{x}$. Note that $\mathcal{Z}^-$ represents all historical observations, and $\mathcal{Z}$ represents all available observations including $\bm{x}_B$, i.e., $p(\bm{x}|\mathcal{Z}) = p(\bm{x}|\{\mathcal{Z}^-,\bm{x}_B\})$. Although the PDR prediction is not explicitly reflected in \eqref{equ:posterior}, it is closely related to $p(\bm{x}|\mathcal{Z}^-)$, which is calculated as:
\begin{equation}
\label{equ:state}
    p(\bm{x}|\mathcal{Z}^-) = \int_\mathcal{M} p(\bm{x}|\bm{x}^-)p(\bm{x}^-|\mathcal{Z}^-)\mathrm{d}\bm{x}^-.
\end{equation}
In \eqref{equ:state}, the PDF of the state transition model, $p(\bm{x}|\bm{x}^-)$, is determined by the PDR prediction.

According to the minimum mean square error (MMSE) principle and constraint \eqref{equ:pf2}, we minimize the MSE between the true position $\bm{x}$ and the estimated position $\widehat{\bm{x}}$ by:
\begin{equation}
\label{equ:pf1}
    \min_{\widehat{\bm{x}}}~\mathbb{E}_{\bm{x}|\mathcal{Z}}[(\bm{x}-\widehat{\bm{x}})^2]~~\mathrm{s.t.}~~\text{\eqref{equ:pf2}}.
\end{equation}
If we neglect constraint \eqref{equ:pf2}, let $\mathrm{d}\mathbb{E}_{\bm{x}|\mathcal{Z}}[(\bm{x}-\widehat{\bm{x}})^2]/\mathrm{d}\widehat{\bm{x}} = 0$ in \eqref{equ:pf1}, and then we can obtain the optimal estimation $\widehat{\bm{x}}^*$ by:
\begin{equation}
\label{equ:target}
    \widehat{\bm{x}}^* = \mathbb{E}_{\bm{x}|\mathcal{Z}}[\bm{x}] = \int_{\mathcal{M}} \bm{x} \cdot p(\bm{x}|\mathcal{Z}) \mathrm{d}\bm{x}.
\end{equation}
However, considering the complexity of PDF $p(\bm{x}|\mathcal{Z})$ and the irregularity of area $\mathcal{W}$, problem \eqref{equ:pf1} does not admit a closed-form solution generally. Although methods such as PF (see Section~\ref{ssec:fuse}) can approximate the numerical solution to \eqref{equ:target}, they still struggle to satisfy constraint \eqref{equ:pf2}. To effectively extract the $\mathcal{W}$ area and incorporate floor plan constraints, we will detail the proposed algorithm in the following sections.

\section{GML Algorithm for BLE Positioning}
\label{sec:ble}
In this section, we propose the GML algorithm based on the MLE principle, to estimate the BLE positioning result $\bm{x}_B$. Assume there are $N'$ beacons distributed in the positioning area, with their map coordinates be $\bm{b}_i\in \mathcal{B}$, $i=1,2,\dots,N'$. The GML algorithm first selects $N\geq3$ beacons with the largest RSSIs for positioning\cite{Bahl2000}. Suppose the set of selected beacons is:
\begin{equation}
    \mathcal{B}_N = \{\bm{b}^{(1)},\bm{b}^{(2)},\dots,\bm{b}^{(N)} \} \subset \mathcal{B},
\end{equation}
where the map coordinate of the $i$-th beacon is $\bm{b}^{(i)} = [x_i, y_i]^\T$. 
Let the RSSI received from $\bm{b}^{(i)}$ after Kalman filtering be denoted as $\widehat{R}_i$. According to \eqref{equ:rssi}, $\widehat{R}_i$ still follows a Gaussian distribution with its variance denoted as $\widehat{\sigma}_i^2$, giving:
\begin{subequations}
\label{equ:rssis}
\begin{align}
    \widehat{R}_i &= -10n\log d_i + R_{0} + X(\widehat{\sigma}_i)\label{equ:rssi1}\\
    &= -10n\log \widehat{d}_i + R_0,\label{equ:rssi2}
\end{align}
\end{subequations}
where $d_i$ and $\widehat{d}_i$ are the true and estimated distances, respectively. From \eqref{equ:rssi1} and \eqref{equ:rssi2}, we obtain: 
\begin{equation}
    \log\widehat{d}_i=\frac{X(\widehat{\sigma}_i)}{10n}+\log d_i \sim \mathcal{N}\left(\log d_i, \frac{\widehat{\sigma}_i^2}{(10n)^2}\right),
\end{equation}
which indicates that $\log\widehat{d}_i$ still follows a Gaussian distribution. If we define
\begin{equation}
\label{equ:def}
\begin{cases}
    \mu_i \triangleq \log d_i,\\
    \eta_i^2 \triangleq \widehat{\sigma}_i^2/(10n)^2,
\end{cases}   
\end{equation}
then $\widehat{d}_i$ follows a Log-Gaussian distribution with base 10 and parameters $(\mu_i,\eta_i^2)$, which can be expressed as:
\begin{equation}
    \widehat{d}_i \sim \mathcal{LN}_{10}(\mu_i, \eta_i^2),
\end{equation}
and the posterior PDF of $\widehat{d}_i$ is:
\begin{equation}
    p(\widehat{d}_i | \bm{y})=\frac{1}{\sqrt{2\pi\eta_i^2}\widehat{d}_i\ln10}\exp\left[-\frac{(\log \widehat{d}_i-\mu_i)^2}{2\eta_i^2}\right].
\end{equation}

After obtaining the PDF $p(\widehat{d}_i | \bm{y})$, the expectation of $\widehat{d}_i$ can be then calculated as:
\begin{equation}
\label{equ:exp1}
    \mathbb{E}_{\widehat{d}_i | \bm{y}}[\widehat{d}_i] = \int_{0}^{+\infty} \widehat{d}_i p(\widehat{d}_i|\bm{y})~\mathrm{d} \widehat{d}_i = 10^{\mu_i+\frac{\eta_i^2}{2}\ln 10}.
\end{equation}
Substituting \eqref{equ:def} into \eqref{equ:exp1}, we obtain:
\begin{equation}
\label{equ:dexp}
    \mathbb{E}_{\widehat{d}_i | \bm{y}}[\widehat{d}_i] = d_i\cdot \exp\left[\frac{1}{2}\left(\frac{\widehat{\sigma}_i\ln10}{10n}\right)^2\right],
\end{equation}
which indicates that the expected estimated distance $\widehat{d}_i$ deviates from the real distance $d_i$ due to the presence of noise. To mitigate this impact, GML algorithm does not directly use $\widehat{d}_i$ as the true distance in calculations. Instead, it considers from the perspective of posterior PDF $p(\widehat{d}_i|\bm{y})$. Let the likelihood function of a point $\bm{y}$ with respect to all beacons in $\mathcal{B}_N$ be: 
\begin{equation}
\label{equ:likeli}
    \ell(\bm{y}) = \prod\limits_{i=1}^{N}p(\widehat{d}_i|\bm{y}).
\end{equation}
Taking the logarithm on both sides of \eqref{equ:likeli} yields:
\begin{equation}
    \ln \ell(\bm{y}) = -\sum\limits_{i=1}^{N}\left[\frac{1}{2\eta_i^2}(\log \widehat{d}_i-\mu_i)^2\right]+c,
\end{equation}
where $c$ is a constant independent of $\bm{y}$. {We denote the basic MLE function as:
\begin{equation}
\label{equ:fmle}
    F_{\text{MLE}}(\bm{y})=\sum_{i=1}^N\frac{1}{\eta^2_i}\left(\log\widehat{d}_i-\log\|\bm{y}-\bm{b}^{(i)}\|\right)^2.
\end{equation}
According to the MLE principle\cite{Feng2014}, the optimal estimation is obtained by minimizing $F_{\text{MLE}}(\bm{y})$. However, such methods suffer from accuracy degradation due to the height difference between beacons and handheld devices. To mitigate this error, we further introduce a penalty term on the basis of \eqref{equ:fmle} to enhance the influence of beacons with stronger RSSI on the estimation. We minimize the following objective function:
\begin{equation}
\label{equ:gml}
    \bm{y}^*=\mathop{\arg\min}_{\bm{y}\in \mathcal{G}'}\left[F_{\text{MLE}}(\bm{y}) + \kappa\sum_{i=1}^N \rho_i\| \bm{y}-\bm{b}^{(i)} \|\right],
\end{equation}
where $\kappa>0$ and $0<\rho_i<1$ are weighting parameters, and $\rho_i$ is defined to increase with the received RSSI $\widehat{R}_i$, which can be computed as follows:
\begin{equation}
    \rho_i=\softmax_{1\leq i\leq N}\tau\widehat{R}_i,
\end{equation}
where $\tau>0$ is a tunable parameter.} Moreover, in \eqref{equ:gml}, $\mathcal{G}'$ is the set of candidate points. Generally, \eqref{equ:gml} is a nonlinear optimization problem that is challenging to solve. We construct a set of grid points distributed throughout the entire area (see Section~\ref{sssec:gac}), denoted as $\mathcal{G}$, and transform \eqref{equ:gml} into a problem of selecting an optimal point within a discrete grid point set $\mathcal{G}' \subset \mathcal{G}$. Let $\mathrm{int}\mathcal{C}_N$ be the interior area of the convex hull $\mathcal{C}_N = \mathcal{CH}(\mathcal{B}_N)$\cite{Pan2024} formed by all beacons in $\mathcal{B}_N$, and $d(\bm{y}_1, \bm{y}_2)$ denote the Manhattan distance between point $\bm{y}_1$ and $\bm{y}_2$. When beacons are densely deployed in the positioning area, we can assume that the user's position is always within $\mathrm{int}\mathcal{C}_N$, then the candidate grid points should also be close to the previous position $\bm{x}^{(k-1)}$:
\begin{equation}
\label{equ:g1}
    \mathcal{G}'_D = \{ \bm{y} \mid \bm{y} \in (\mathcal{G}\cap\mathrm{int}\mathcal{C}_N) \wedge d(\bm{y}, \bm{x}^{(k-1)}) < d_{0}\}.
\end{equation}
In this way, we can effectively reduce the time delay caused by searching through a large number of grid points in $\mathcal{G}$.

{If the beacons are deployed sparsely, i.e., $\mathrm{int}\mathcal{CH}(\mathcal{B})$ cannot geometrically cover the entire area, $\mathrm{int}\mathcal{C}_N$ may fail to cover the actual user position, and thus may instead lead to positioning deviation. Therefore, we remove the convex hull constraint $\mathrm{int}\mathcal{C}_N$ and revise \eqref{equ:g1} accordingly:}
\begin{equation}
\label{equ:g2}
    \mathcal{G}'_S = \{ \bm{y} \mid \bm{y} \in \mathcal{G} \wedge d(\bm{y}, \bm{x}^{(k-1)}) < d_{0}\}.
\end{equation}
In the experimental scenario of this paper, 
{for simplicity, it is assumed that the $\eta_i$ of all beacons are equal\cite{Feng2014} and normalized as 1.} The BLE positioning result at current time $t$ is obtained by averaging $\bm{y}^*$ and the previous $(n-1)$ BLE positioning results: 
\begin{equation}
    \bm{x}_B^{(t)} = \frac{1}{n}\left(\bm{y}^*+\sum_{i=1}^{n-1}\bm{x}_B^{(t-i)}\right).
\end{equation}

{From the above analysis, it is evident that GML algorithm takes the RSSI noise into consideration, uses the posterior PDF of $\widehat{d}_i$ to form a unique likelihood, and introduces a novel penalty term, thereby reducing errors.} To fuse BLE results into our positioning system, we will detail the proposed FP-BP algorithm in the following section.

\section{FP-BP Algorithm}
\label{sec:alg}

This section proposes the FP-BP algorithm, which consists of four steps: (i) preprocessing the floor plan, (ii) obtaining the PDR result, (iii) fusing BLE with PDR, and (iv) performing necessary PPC based on the floor plan.

\subsection{Floor Plan Preprocessing}\label{ssec:fpp}
To deeply integrate floor plans into the positioning system, preprocessing should be carried out in advance. The main steps of preprocessing include stylization, feature extraction, and map construction.

\begin{figure}[!t]
\begin{subfigure}{0.2\textwidth}
    \centering
    \includegraphics[height=0.575\textwidth]{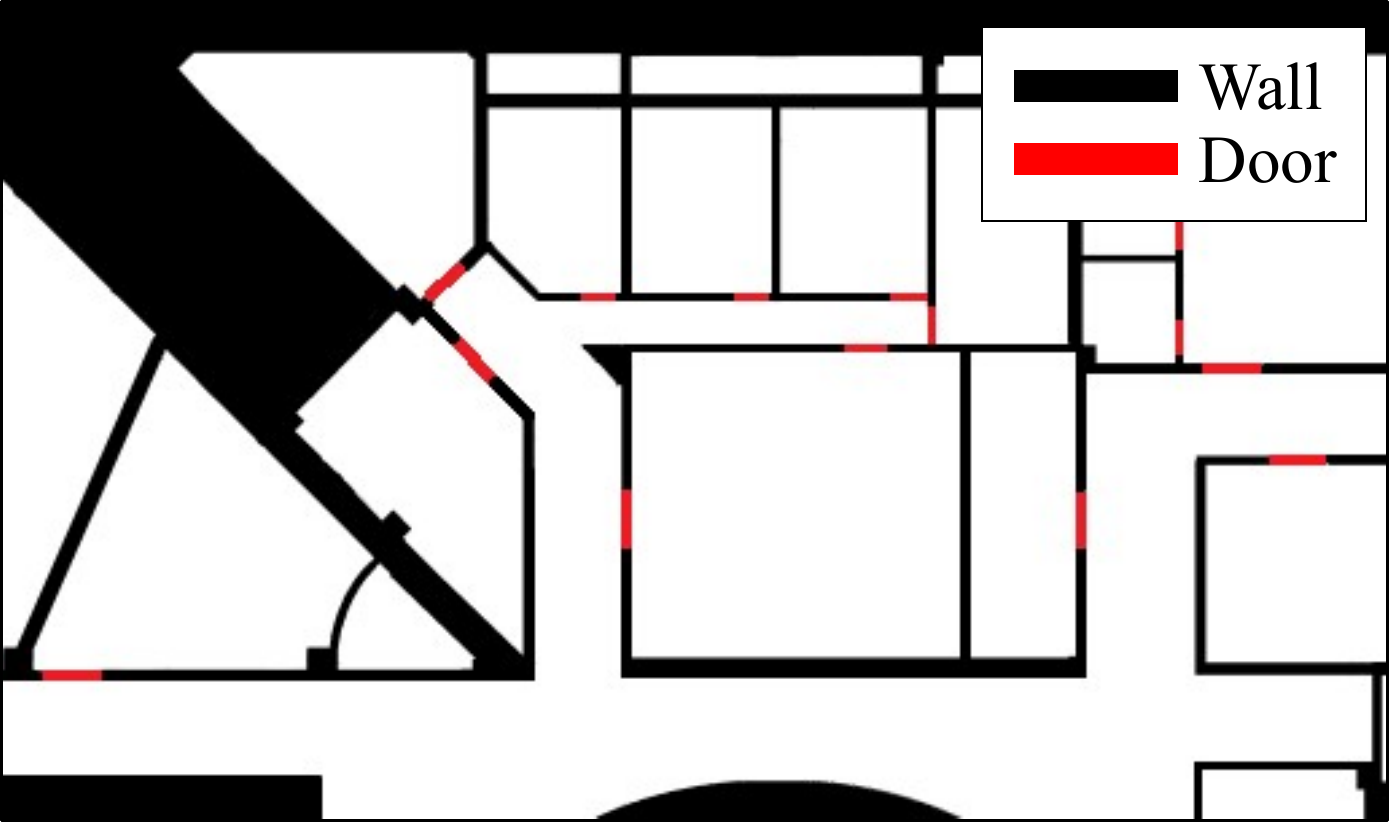}
    \caption{Part of the stylized floor plan.}
    \label{fig:par_a}
\end{subfigure}
\hspace{2pt}
\begin{subfigure}{0.2\textwidth}
    \centering
    \includegraphics[height=0.575\textwidth]{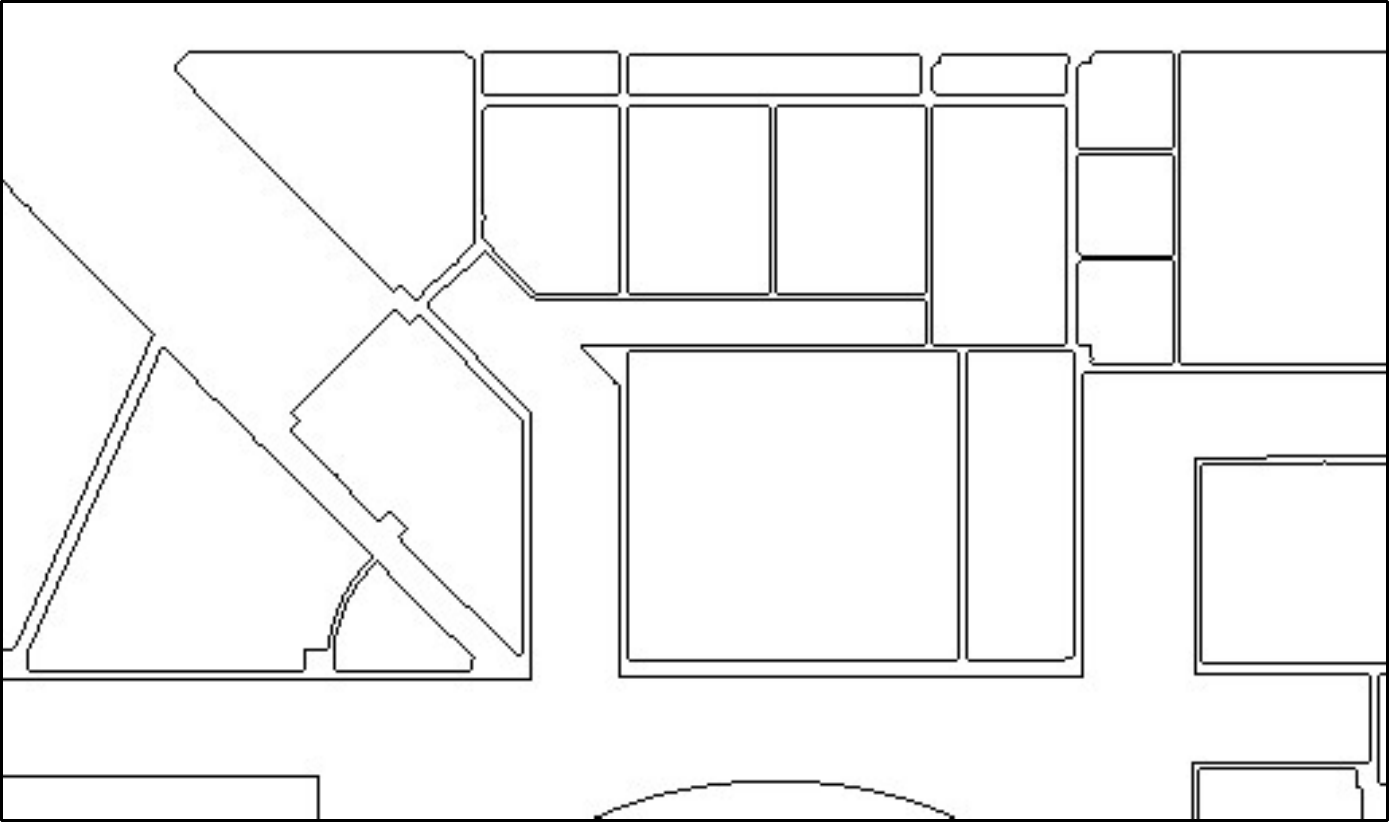}
    \caption{Part of the inner contours.}
    \label{fig:par_b}
\end{subfigure}
\centering

\begin{subfigure}{0.45\textwidth}
    \centering
    \includegraphics[width=\textwidth]{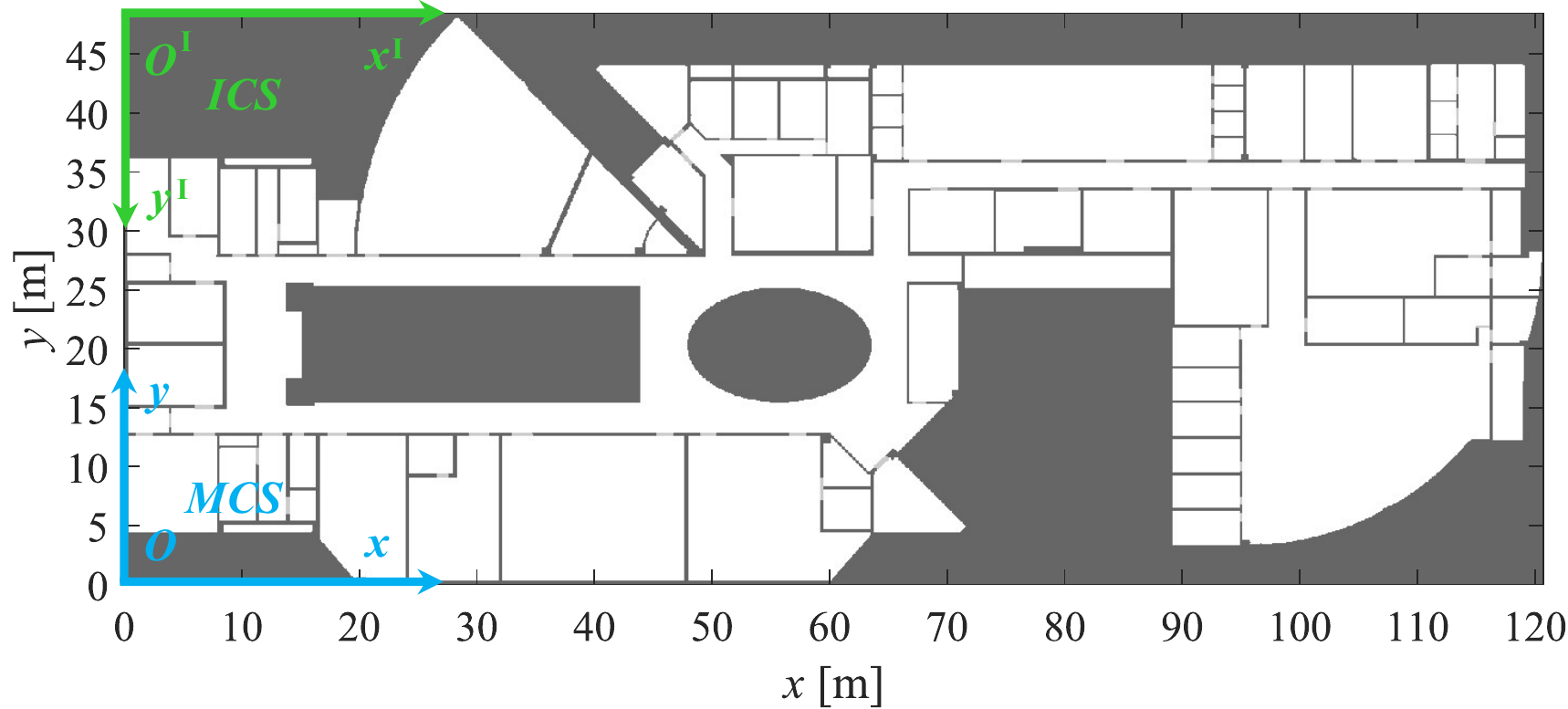}
    \caption{Processed map $\bm{M}(\bm{x})$.}
    \label{fig:par_c}
\end{subfigure}
\centering
\caption{Offline preprocessing of the experimental area floor plan.}
\label{fig:par}
\end{figure}

\subsubsection{Stylization}
{To eliminate the distractions of computer-aided design (CAD) drawings, we manually redraw a stylized floor plan for positioning.} In particular, the floor plan map features required by the algorithm often include \textit{Wall} (i.e., non-passable obstacles) and \textit{Door} (i.e., conditionally passable obstacles). We use distinguishable colors to fill the \textit{Wall} and \textit{Door} areas, such as black and red, while the \emph{Walkable} areas are filled with white, as illustrated in Fig.~\ref{fig:par}\subref{fig:par_a}. 

\subsubsection{Feature Extraction}
\label{sssec:fe}
To identify the \emph{Rooms} and obtain the precise areas of different map features, we need to extract these features from the floor plan. A mask-based method simplifies this process effectively. Adding the binary masks\cite{Vincent1993} of \emph{Walls}, $\bm{I}_{\textit{Wall}}$, and \emph{Doors}, $\bm{I}_{\textit{Door}}$, we obtain the mask of the whole obstacle area $\bm{I}_{\textit{obs}}$ by $\bm{I}_{\textit{obs}} = \bm{I}_{\textit{Wall}} + \bm{I}_{\textit{Door}}$. Then, we extract the inner contours of $\bm{I}_{\textit{obs}}$, which represent either a \textit{Room} or an obstacle boundary within the \textit{Room}, as shown in Fig.~\ref{fig:par}\subref{fig:par_b}. {Note that this step simply determines the map feature of each pixel individually based on its specific color, which is independent of the overall feature shape, and thus can avoid misjudgment.}

\subsubsection{Map Construction}
\label{sssec:mc}
To efficiently integrate the floor plan into the positioning algorithm, two coordinate systems are defined on the floor plan image: the map coordinate system (MCS) and the image coordinate system (ICS), as shown in Fig.~\ref{fig:par}\subref{fig:par_c}. The transformation between ICS and MCS is:
\begin{equation}
    \begin{bmatrix} x\\y \end{bmatrix} = \frac{1}{r}\begin{bmatrix} x^\mathrm{I}\\H^\mathrm{I}-y^\mathrm{I} \end{bmatrix},
\end{equation}
where $\bm{x} = [x,y]^\T$ and $\bm{x}^\mathrm{I} = [x^\mathrm{I},y^\mathrm{I}]^\T$ are the map coordinate and image coordinate respectively; $r$ is the resolution factor that represents the number of pixels in ICS corresponding to a distance of 1 m in MCS, and $H^\mathrm{I}$ is the height of the floor plan image in pixels. 

For a given point in MCS, we need to obtain its map feature to determine which area the point is in, including \emph{Walkable}, \emph{Wall}, and \emph{Door} areas. This allows us to correct the coordinates of this point based on its map feature. Based on the MCS, we define a map feature function $\bm{M}(\bm{x}) = \bm{I}(\bm{x}^\mathrm{I}) \in [0,1]$, where:
\begin{equation}\label{equ:mask}
    \bm{I}(\bm{x}^\mathrm{I}) = \bm{I}_{\textit{Wall}}(\bm{x}^\mathrm{I}) + 0.5\bm{I}_{\textit{Door}}(\bm{x}^\mathrm{I}).
\end{equation}
For a point $\bm{x}$ in the MCS, querying $\bm{M}(\bm{x})$ determines the map feature to which the point $\bm{x}$ belongs:
\begin{equation}
    \bm{M}(\bm{x}) = \begin{cases}
        0,&\bm{x}~\textit{is \textbf{Walkable}},\\
        0.5,&\bm{x}~\textit{is \textbf{Door}},\\
        1,&\bm{x}~\textit{is \textbf{Wall}}.
    \end{cases}
    \label{equ:map}
\end{equation}
The role of the floor plan is to restrict the estimated position $\bm{x}$ within the \emph{Walkable} area $\mathcal{W} = \{\bm{x}\mid\bm{M}(\bm{x}) = 0\}$, since it is generally unrealistic for pedestrians to walk or remain in obstacle areas. Note that in the following text, we will abbreviate this map feature function as the \emph{map}.

\subsubsection{Grid Array Construction}
\label{sssec:gac}
In Section~\ref{sec:ble}, we have proposed the GML algorithm, which selects the grid point with the maximum likelihood probability as the positioning result. To apply the GML algorithm for BLE positioning, we need to construct a virtual grid array in advance, where the grid points are distributed over the entire map. Also, we should remove all the grid points within obstacle areas. Let the map range be $[0,X]\times[0,Y]$, and the grid interval be $I_M$. Then, the position of the $(i,j)$-th grid point is $\bm{g}_{ij}=[(i+0.5)X/I_M,(j+0.5)Y/I_M]^\T$, where $1\leq i \leq \lfloor X/I_M \rfloor-1$, $1\leq j \leq \lfloor Y/I_M \rfloor-1$, and $\lfloor\cdot\rfloor$ denotes the floor function. Let the grid array be denoted as $\mathcal{G}$, and then all grid points in $\mathcal{G}$ are located in \textit{Walkable} area, i.e., $ \mathcal{G}=\{ \bm{g}_{ij} \mid \bm{M}(\bm{g}_{ij})=0 \}$. Substituting this grid array $\mathcal{G}$ into \eqref{equ:g1} and \eqref{equ:g2}, we can ensure that the BLE estimation result is located solely within \emph{Walkable} area.

All of the above steps can be completed during the offline phase. Additionally, we note that: (i) When the floor plan is too large, downsampling can be applied before \textit{Stylization} to avoid an excessively high map resolution factor $r$, which would otherwise increase computational cost. {(ii) If the building structure changes, only \textit{Stylization} requires manual adjustment, while the remaining steps are fully automated, thereby minimizing system update and maintenance costs.}

\subsection{Pedestrian Dead Reckoning (PDR)}
\label{ssec:pdr}

\begin{figure}[!t]
\begin{subfigure}{0.19\textwidth}
    \includegraphics[height=\textwidth]{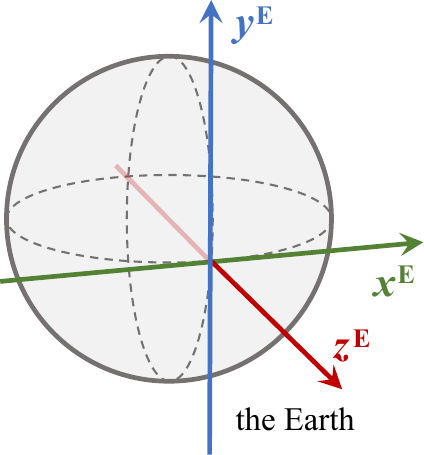}
    \centering
    \caption{The ECS.}
    \label{fig:ecs}
\end{subfigure}
\begin{subfigure}{0.19\textwidth}
    \centering
    \includegraphics[height=\textwidth]{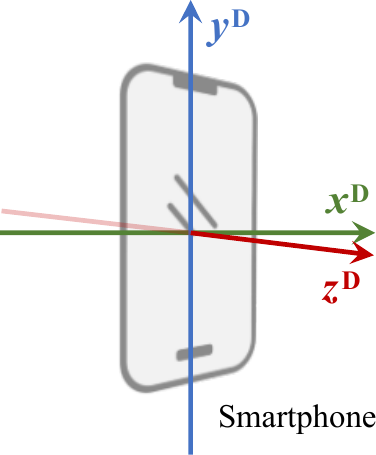}
    \caption{The DCS.}
    \label{fig:dcs}
\end{subfigure}
\centering
\caption{The defined 3-D ECS and DCS\cite{Android}.}
\label{fig:css}
\end{figure}

In this subsection, the following coordinate systems are first defined: (i) the east-north-up coordinate system (ECS), whose axes are parallel to the east, north, and upward directions, as shown in Fig.~\ref{fig:css}\subref{fig:ecs}; (ii) the device coordinate system (DCS), which is defined by the Android Developers\cite{Android} as shown in Fig.~\ref{fig:css}\subref{fig:dcs}; (iii) the 3-dimensional (3-D) MCS, where the $xy$-plane is the same as that of the 2-D MCS in Section~\ref{sssec:mc}, and the $z$-axis is determined by the cross product of the $x$- and $y$-axes. In reality, the $z$-axis of 3-D MCS is considered aligned with that of ECS.

The rotation from DCS to ECS is typically obtained in the form of quaternions from the gyroscope, denoted as $\bm{q}_\mathrm{D}^\mathrm{E} = q_x\mathrm{i}+q_y\mathrm{j}+q_z\mathrm{k}+q_w$. By converting $\bm{q}_\mathrm{D}^\mathrm{E}$ to a rotation matrix $\bm{R}_\mathrm{D}^\mathrm{E}$, we obtain:
\begin{equation}
    \resizebox{0.89\hsize}{!}{$
    \bm{R}_\mathrm{D}^\mathrm{E} = 
    \begin{bmatrix}
        1-2q_y^2-2q_z^2 & 2q_xq_y-2q_zq_w & 2q_xq_z+2q_yq_w \\
        2q_xq_y+2q_zq_w & 1-2q_x^2-2q_z^2 & 2q_yq_z-2q_xq_w \\
        2q_xq_z-2q_yq_w & 2q_yq_z+2q_xq_w & 1-2q_x^2-2q_y^2 \\
    \end{bmatrix}.
    $}
\end{equation}
The rotation matrix from ECS to 3-D MCS, $\bm{R}_\mathrm{E}$, is constant, and the rotation matrix from DCS to 3-D MCS is given by:
\begin{equation}
    \bm{R}_{\mathrm{D}} = \bm{R}_\mathrm{E}\cdot\bm{R}_{\mathrm{D}}^\mathrm{E}.
\end{equation}

After defining these coordinate systems, it is essential to accurately detect the step events, and calculate the step lengths and yaws in the PDR system. The step detection relies on the smartphone acceleration $\bm{a}$ in 3-D MCS, but the linear acceleration $\bm{a}^{\mathrm{D}}$ obtained from the accelerometer is in DCS. We calculate the acceleration $\bm{a}$ by:
\begin{equation}
    \bm{a} = \bm{R}_\mathrm{D} \cdot \bm{a}^{\mathrm{D}}.
\end{equation}
Let the current time be $k$, and the sequence of the vertical component of $\bm{a}$ be $\{ z_i \}_{i=1}^k$. The proposed conditions for determining the step event are as follows:
\begin{subequations}
\label{equ:app}
\begin{numcases}{}
    z_{k-h} = \max\limits_{i = k-2h}^{k} z_i,\\
    z_{k-h} > z_{\mathrm{th}},\\
    k-h-h'>k_{\mathrm{th}},
\end{numcases}
\end{subequations}
where $h$ is half of the window size; $z_{\mathrm{th}}$ is the threshold for maximum step acceleration; $k_{\mathrm{th}}$ is the threshold for minimum step interval; and $h'$ is the last step event time. Only when these three conditions are simultaneously satisfied can time $k-h$ be recognized as a step event. At this time, the peak-to-peak value of the acceleration within this $2h$ window is calculated as:
\begin{equation}
    z_{\mathrm{pp}}=z_{k-h}-\min_{i = k-2h}^{k} z_i,
\end{equation}
and the step length $s$ is estimated as\cite{weinberg2002}:
\begin{equation}
\label{equ:step}
    s = \beta \sqrt[4]{z_{\mathrm{pp}}}.
\end{equation}
The step direction vector in 3-D MCS is estimated as:
\begin{equation}
\label{equ:dir3d}
    \bm{d}_k' = 
    \begin{cases}
        \bm{R}_\mathrm{D}[0,1,0]^\T, &k \leq k_0, \\
        \bm{R}_\mathrm{M}(\bm{\omega}^\mathrm{D},\bm{a}_g^\mathrm{D},\bm{m}^\mathrm{D}; \theta_{k_0})\bm{R}_\mathrm{D}[0,1,0]^\T, &k>k_0,
    \end{cases}
\end{equation}
where $\bm{R}_\mathrm{M}(~\cdot~;\theta_{k_0})$ represents the orientation filter proposed by Madgwick et al.\cite{Madgwick2011} with initial yaw $\theta_{k_0}$. This filtering algorithm takes the angular acceleration $\bm{\omega}^\mathrm{D}$, acceleration $\bm{a}_g^\mathrm{D}$ that includes gravity, and magnetic flux density $\bm{m}^\mathrm{D}$ as inputs. The yaw at time $k$ is then estimated as:
\begin{equation}
    \theta_k = \mathrm{atan2}(\bm{d}_k),
\end{equation}
where $\bm{d}_k$ is the normalized projection of $\bm{d}_k'$ from 3-D MCS to 2-D MCS. In this way, we obtain the step length $s_k$ and yaw $\theta_k$ if a step event is detected at time $k$.

\subsection{Fusion with PF}
\label{ssec:fuse}
We use the PF to fuse the BLE result in Section~\ref{sec:ble} and the PDR result in Section~\ref{ssec:pdr}. A PF at time $k$ can be defined as $\mathcal{P}^{(k)} = \{ \bm{x}_i^{(k)}, w_i^{(k)} \}_{i = 1}^m$, where $\bm{x}_i^{(k)}$ is the position of $i$-th particle, and $w_i^{(k)}$ is its weight. When $k = 0$, we set the initial state as $\bm{x}_i^{(0)} = \overline{\bm{x}}_B^{(0)}$, $w_i^{(0)} = 1/m$, where $\overline{\bm{x}}_B^{(0)}$ is the average of BLE positioning results during initialization. 

The state transition of the $i$-th particle can be modeled as:
\begin{equation}
    \widehat{\bm{x}}^{(k)}_i = \bm{x}_i^{(k-1)} + (s_{k-1}+\nu_{s,i}) \cdot \begin{bmatrix}
        \cos (\theta_{k-1} + \nu_{\theta,i}) \\
        \sin (\theta_{k-1} + \nu_{\theta,i})
    \end{bmatrix},
\end{equation}
where $\nu_{s,i}$ and $\nu_{\theta,i}$ are the zero-mean PDR noise, and are modeled as the uniform distribution. In the PF, we use the Monte Carlo method to approximate \eqref{equ:posterior} and \eqref{equ:target} by:
\begin{subequations}
\label{equ:mc}
\begin{numcases}{}
    p(\bm{x}^{(k)}|\mathcal{Z}^{(k)}) \approx \textstyle\sum\limits_{i=1}^m \widetilde{w}_i^{(k)} \delta_D(\bm{x}^{(k)}-\widehat{\bm{x}}^{(k)}_i),\label{equ:mc1}\\
    \widehat{\bm{x}} \approx \textstyle\sum\limits_{i=1}^m \widetilde{w}_i^{(k)} \widehat{\bm{x}}^{(k)}_i, \label{equ:mc2}
\end{numcases}
\end{subequations}
where $\delta_D(\cdot)$ is the Dirac function. According to \eqref{equ:posterior}, the normalized particle weights $\{\widetilde{w}_i^{(k)}\}_{i=1}^m$ should be calculated as normalized $p(\bm{x}_B^{(k)}|\bm{x}_i^{(k)})$. Therefore, based on the assumption of Gaussian noise\cite{Xu2019}, these weights are calculated by:
\begin{equation}
    w_i^{(k)} = \frac{1}{\sqrt{2\pi \sigma_x^2}}\exp{\left[-\frac{\| \bm{x}_B^{(k)} - \widehat{\bm{x}}_i^{(k)}\|^2}{2\sigma_x^2}\right]},
\end{equation}
and then be normalized. Finally, by resampling the PF\cite{Douc2005}, the estimated position $\widehat{\bm{x}}$ is calculated as the average of all particles. As mentioned before, $\widehat{\bm{x}}$ is a numerical approximation of \eqref{equ:target} and does not take constraint \eqref{equ:pf2} into account. We will address this problem in the following subsection.

\begin{algorithm}[!t]
    \renewcommand{\algorithmcfname}{Function}
    \caption{\textit{Raycast}}
    \small
    \label{alg:Raycast}
    \KwIn{Map: $\bm{M}(\bm{x})$;~Start point: $\bm{x}_1$;~End point: $\bm{x}_2$\;
        \hspace{3em}{[Update step: $\delta<r^{-1}$];}
    }
    \KwOut{Hit flag: \texttt{hit};~Hit point: $\bm{h}$\;}
    $\texttt{hit} = \mathbb{F}$\;
    \If{$\bm{x}_1 - \bm{x}_2 = \mathbf{0}$}
    {   
        \textbf{return} $\texttt{hit} = (\bm{M}(\bm{x}_1) > 0)$, $\bm{h} = \texttt{hit}~\textbf{?}~\bm{x}_1~\textbf{:}~\mathbf{Null}$\;
    }
    $N = \mathrm{Round}(\|\bm{x}_2-\bm{x}_1\|/\delta)$\;
    $\bm{x}_\mathrm{n} = (\bm{x}_2-\bm{x}_1)/\|\bm{x}_2-\bm{x}_1\|$\;
    \For{$k = 0:N$}
    {
        \If{$\bm{M}(\bm{x}_1 + k\delta\bm{x}_\mathrm{n}) > 0$}
        {
            \textbf{return} $\texttt{hit} = \mathbb{T}$, $\bm{h} = \bm{x}_1 + k\delta\bm{x}_\mathrm{n}$\;
        }
    }
    \lIf{$\text{!}~\texttt{hit}$}{\textbf{return} $\texttt{hit} = \mathbb{F}$, $\bm{h} = \mathbf{Null}$}
\end{algorithm}

\subsection{Floor Plan Assisted Post Processing}

To integrate floor plan constraint \eqref{equ:pf2} into our system, we propose the PPC algorithm, which consists of two parts: \emph{Correction Determination} and \emph{Position Correction}. The former is responsible for deciding whether the initial estimation $\widehat{\bm{x}}$ in Section~\ref{ssec:fuse} needs correction, and the latter applies specific corrections to the estimated position based on different determination conditions.

\subsubsection{Correction Determination}
\label{sssec:cd}
We first define the \emph{Raycast} function, which is summarized in \textbf{Function \ref{alg:Raycast}}. In \emph{Raycast}, we move from a start point incrementally towards the end point, until it is reached. If an obstacle is encountered during this process, the \emph{Raycast} returns $\mathbb{T}$ and the first contact point $\bm{h}$ (also referred to as the \emph{hit point}). \emph{Raycast} efficiently detects whether there is an obstacle between two points; if an obstacle exists, it can find the hit point with a time complexity of $\mathcal{O}(n)$. {Note that the update step $\delta$ should be set such that $\delta<r^{-1}$ to prevent missed detections, where $r$ is the map resolution, since any obstacle thinner than one pixel (i.e., $r^{-1}$ in reality) would otherwise be ignored in the image.}

At a time $k$, perform \emph{Raycast} from the previous position $\bm{x}^{(k-1)}$, to the estimated position $\widehat{\bm{x}}$. If the hit point $\bm{h}$ is obtained, it is considered that moving from $\bm{x}^{(k-1)}$ to $\widehat{\bm{x}}$ is not feasible. The infeasible step necessitates a position correction, which typically includes step length correction and yaw correction. Then, a candidate rotation angle sequence $\mathbf{\Phi} = \{ \pm i\cdot\Delta\phi \}_{i=1}^{N}$ is defined. In this sequence, $\Delta\phi$ is a unit rotation angle, and the elements in $\mathbf{\Phi}$ are arranged in ascending order of their absolute values.

From Section~\ref{sssec:mc}, we can determine the map feature $\bm{M}(\bm{h})$ of the hit point $\bm{h}$:

(i) If $\bm{h}$ is \emph{Wall}, i.e., $\bm{\bm{M}}(\bm{h})=1$, it is assumed to correct yaw by searching both sides for \emph{Walkable} areas, i.e., to find the best angle by directly rotating in the order of $\mathbf{\Phi}$.

(ii) If $\bm{h}$ is \emph{Door}, i.e., $\bm{\bm{M}}(\bm{h}) = 0.5$, it is necessary to determine whether the user should pass through the \emph{Door}, since \emph{Doors} are defined as conditionally passable obstacles. If it is determined that the user should not pass through the \emph{Door}, the PPC algorithm assumes that the yaw should be corrected towards the inside of the \emph{Room}.

We use the incident angle $\alpha$ of the estimated step vector $\widehat{\bm{s}}$ to determine whether the user passes through the \emph{Door}. Let $\widehat{\bm{s}} = \widehat{\bm{x}} - \bm{x}^{(k-1)}$, and the incident angle $\alpha$ of $\widehat{\bm{s}}$ to the obstacle can then be calculated as: 
\begin{equation}
\label{equ:inc}
    \alpha = \langle \bm{n}_h, -\widehat{\bm{s}} \rangle,
\end{equation}
where $\bm{n}_h$ is the normal vector at $\bm{h}$, obtained by transforming the image gradient vector $\nabla \bm{I}(\bm{h}^\mathrm{I})$ from ICS to MCS; $\langle\bm{x},\bm{y}\rangle$ is the included angle formed by vectors $\bm{x}$ and $\bm{y}$. When $\alpha$ is larger than a threshold $\alpha_0$, it is assumed that the user should not pass through the \emph{Door}, and in this case, we rotate the estimated step $\widehat{\bm{s}}$ by a positive and a negative angle, $\phi_+$ and $\phi_-$, respectively. Then the resulting incident angles, $\alpha_+$ and $\alpha_-$, are compared. According to the geometric principle, rotating towards the inside of \emph{Room} is equivalent to rotating in the direction of increasing incident angles. This allows the determination of the sign of the correction angle, i.e., the direction of rotation, and $\mathbf{\Phi}$ is filtered to retain only the elements corresponding to the determined sign, while maintaining their original order.

\begin{figure}[!t]
\begin{subfigure}{0.21\textwidth}
    \includegraphics[height=\textwidth]{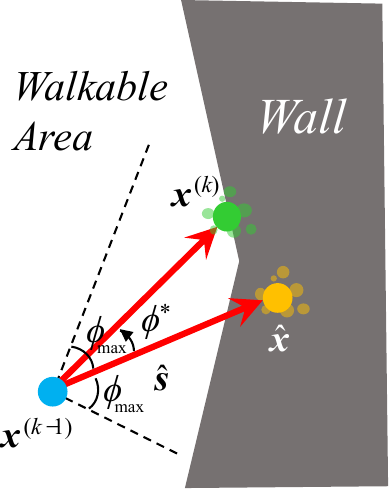}
    \centering
    \caption{}
    \label{fig:case1}
\end{subfigure}
\begin{subfigure}{0.21\textwidth}
    \centering
    \includegraphics[height=\textwidth]{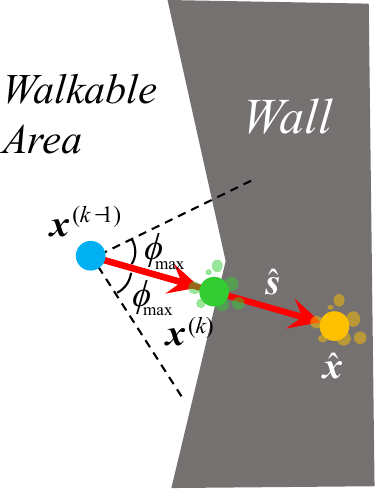}
    \caption{}
    \label{fig:case2}
\end{subfigure}
\centering
\caption{Position correction part I: yaw correction is needed. In (a), the step is rotated by an optimal angle $\phi^*$ towards the \textit{Walkable} area (Case 1); while in (b), the step is considered blocked (Case 2).}
\label{fig:cases1}
\end{figure}

\subsubsection{Position Correction}

In Section~\ref{sssec:cd}, we proposed the method for determining whether the yaw should be corrected, and it was concluded that in PPC algorithm, the only case where yaw correction is not required is when the pedestrian passes through a \emph{Door}.

(i) If the yaw correction is needed, as shown in Fig.~\ref{fig:cases1}\subref{fig:case1}, we find the optimal correction rotation angle by sequentially traversing all candidate angles in $\mathbf{\Phi}$ and then determining the optimal corrected position. $\forall\phi \in \mathbf{\Phi}$, we can construct a rotation matrix by:
\begin{equation}
    \bm{R}_\phi \triangleq \begin{bmatrix}
        \cos \phi & -\sin \phi \\
        \sin \phi & \cos \phi
    \end{bmatrix},
\end{equation}
and rotate $\widehat{\bm{s}}$ to obtain a test point $\bm{t}_\phi$:
\begin{equation}
\label{equ:tp}
    \bm{t}_\phi = \bm{x}^{(k-1)} + \bm{R}_\phi \cdot f\widehat{\bm{s}},
\end{equation}
where $f \geq 1$ is a scale factor. Next, we follow the order of $\mathbf{\Phi}$, sequentially searching for the optimal rotation angle $\phi^*$ with the smallest absolute value, such that \emph{Raycast} from $\bm{x}^{(k-1)}$ to $\bm{t}_\phi$ returns $\mathbb{F}$, which indicates that this path is feasible without any obstruction:
\begin{subequations}\label{equ:optpih}
\begin{gather}
    \phi^* = \arg \min_{\phi\in\mathbf{\Phi}}\,|\phi| \tag{\theequation}\label{equ:optpih1}\\
    \mathrm{s.t.}~~\mathrm{Raycast}(\bm{M}, \bm{x}^{(k-1)}, \bm{t}_\phi) = \mathbb{F}.\label{equ:optpih2}
\end{gather}
\end{subequations}
Then, the corrected position defined as: 
\begin{equation}
\label{equ:case1}
    \bm{x}^{(k)} = \bm{x}^{(k-1)} + \bm{R}_{\phi^*} \cdot \widehat{\bm{s}}.
\end{equation}

If a suitable rotation angle cannot be found within sequence $\mathbf{\Phi}$, i.e., $\forall\phi\in\mathbf{\Phi},~\mathrm{Raycast}(\bm{M},\bm{x}^{(k-1)},\bm{t}_\phi)=\mathbb{T}$, we can only assume that the step is blocked by an obstacle. As shown in Fig.~\ref{fig:cases1}\subref{fig:case2}, in this case, a step length correction is applied instead, by moving a small distance backward from the hit point $\bm{h}$ to \emph{Walkable} area, yielding the corrected position: 
\begin{equation}
\label{equ:case2}
    \bm{x}^{(k)} = \bm{h} - \varepsilon \widehat{\bm{s}}.
\end{equation}

\begin{figure}[!t]
\begin{subfigure}{0.21\textwidth}
    \includegraphics[height=\textwidth]{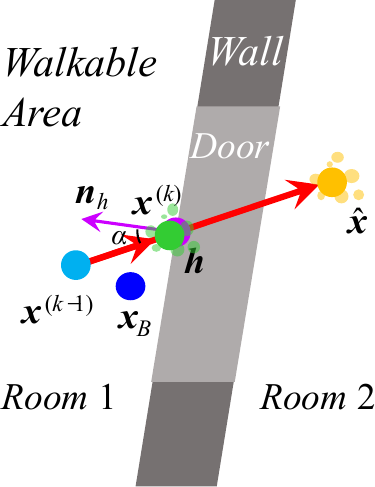}
    \centering
    \caption{}
    \label{fig:case2-1}
\end{subfigure}
\begin{subfigure}{0.21\textwidth}
    \centering
    \includegraphics[height=\textwidth]{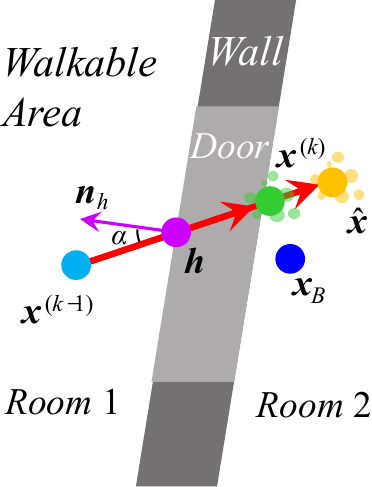}
    \caption{}
    \label{fig:case3}
\end{subfigure}
\centering
\caption{Position correction part II: yaw correction is unnecessary. In (a), the step is considered blocked by the \textit{Door} (Case 2); while in (b), the pedestrian is considered to pass through the \textit{Door} (Case 3).}
\label{fig:cases2}
\end{figure}

\begin{figure}[!t]
\centering
\includegraphics[width=0.42\textwidth]{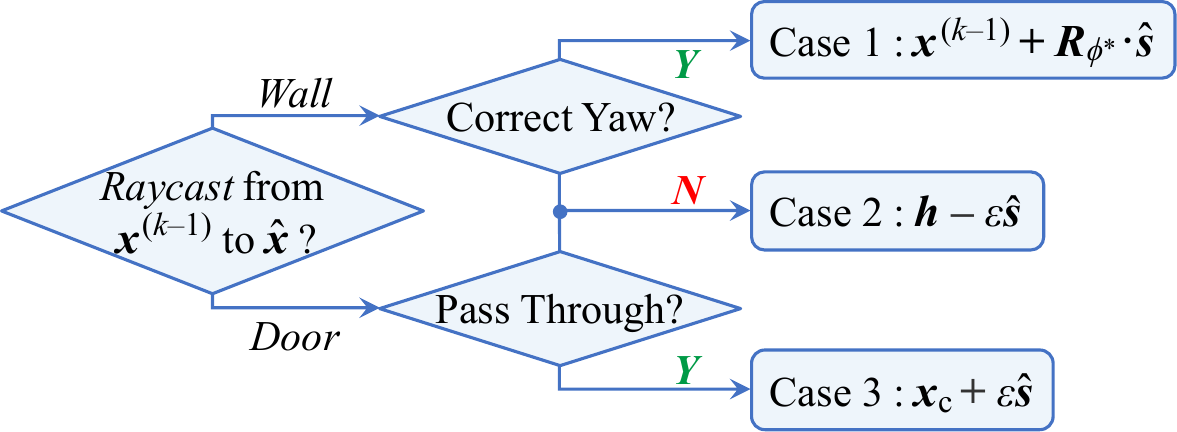}
\caption{Flow chart of position correction in PPC algorithm.}
\label{fig:ppc}
\end{figure}

(ii) If the yaw correction is unnecessary, even though the incident angle suggests that the user should pass through the \emph{Door}, it is essential to first determine whether the user's current \emph{Room} has changed, since passing through a \emph{Door} always implies a \emph{Room} change. If the \emph{Room} remains unchanged, the user should not be permitted to pass through the \emph{Door}.

\begin{algorithm}[!t]
    \caption{PPC Algorithm}
    \small
    \label{alg:PPC}
    \KwIn{Previous position: $\bm{x}^{(k-1)}$;~Estimated step: $\widehat{\bm{s}}$\;
        \hspace{3em}Map: $\bm{M}(\bm{x})$;~Estimated particles: $\{\widehat{\bm{x}}^{(k)}_i\}_{i=1}^m$\;}
    \KwOut{Estimated position: $\bm{x}^{(k)}$;~~Particles: $\mathcal{P}^{(k)}$\;}
    Set candidate angles as $\mathbf{\Phi}= [\pm j\cdot\Delta\phi~\textbf{for}~j=1, 2, \dots, N]$\;
    Obtain flag \texttt{hit} and point $\bm{h}$ by $\mathrm{Raycast}(\bm{M}, \bm{x}^{(k-1)}, \widehat{\bm{x}})$\;
    \If{\texttt{hit} \textbf{and} $\bm{h} \neq \bm{x}^{(k-1)}$}
    {
        \eIf{$\bm{h}$ is \textbf{Door}}
        {
            Obtain $\bm{h}^\mathrm{I}$ by transforming $\bm{h}$ into ICS\;
            Calculate the image gradient vector $\nabla \bm{I}(\bm{h}^\mathrm{I})$\;
            Obtain $\bm{n}_h$ by transforming $\nabla \bm{I}(\bm{h}^\mathrm{I})$ into MCS\;
            Obtain the incident angle $\alpha$ by \eqref{equ:inc}\;
            \If{$\alpha > \alpha_0$} 
            {
                $\alpha_+ = \langle \bm{n}_h, -\bm{R}_{\phi_+} \cdot \widehat{\bm{s}}\rangle$\;
                $\alpha_- = \langle \bm{n}_h, -\bm{R}_{\phi_-} \cdot \widehat{\bm{s}}\rangle$\;
                $\mathbf{\Phi} = (\alpha_+ > \alpha_-)~\textbf{?}~\mathbf{\Phi}[\mathbf{\Phi} > 0]~\textbf{:}~\mathbf{\Phi}[\mathbf{\Phi} < 0]$\;
            }
            $\texttt{correct\_yaw} = (\alpha > \alpha_0)$\;   
        }       
        { $\texttt{correct\_yaw} = \mathbb{T}$\; }
        \eIf{$\texttt{correct\_yaw}$}
        {
            Find the optimal rotation angle $\phi^*\in\mathbf{\Phi}$ by \eqref{equ:optpih}\;
            \eIf{$\phi^*$ is found}
            { 
                Case 1: Correct the position $\bm{x}^{(k)}$ by \eqref{equ:case1}\;
            }
            { 
                Case 2: Correct the position $\bm{x}^{(k)}$ by \eqref{equ:case2}\;
            }
        }
        {
            Check the \textit{Room} in which $\bm{x}_B$ is\;
            Case 2/3: Correct the position $\bm{x}^{(k)}$ by \eqref{equ:case23}\;
        }
        Update the particles by \eqref{equ:updatepf}\;
    }
\end{algorithm}

Since all the internal contours of the map are obtained in the offline phase in Section~\ref{sssec:fe}, it is easy to determine the \emph{Room} of the last result $\bm{x}^{(k-1)}$, i.e., the contour that contains $\bm{x}^{(k-1)}$. Since the estimated results of GML algorithm are typically accurate enough to distinguish different \emph{Rooms}, we use $\bm{x}_B$ as the basis for determining the current \emph{Room}. As shown in Fig.~\ref{fig:cases2}\subref{fig:case2-1}, if $\bm{x}_B$ is still in the previous \emph{Room}, we still consider it to be blocked by the \emph{Door}, and thus, the result is corrected by moving from the hit point backward to the \emph{Walkable} area. Otherwise, in Fig.~\ref{fig:cases2}\subref{fig:case3}, we consider the user to pass through the \emph{Door} in this step. We identify the closest point $\bm{x}_{\mathrm{c}}$ from all the contours, and correct the result by moving a small step forward from $\bm{x}_{\mathrm{c}}$ into \emph{Walkable} area, completing the action of passing through the \emph{Door}:
\begin{equation}
\label{equ:case23}
    \bm{x}^{(k)} = \begin{cases}
        \bm{h} - \varepsilon \widehat{\bm{s}}, &\textit{if}~\bm{x}_B~\textit{is in previous Room},\\
        \bm{x}_{\mathrm{c}} + \varepsilon \widehat{\bm{s}}, &\textit{otherwise}.
    \end{cases}
\end{equation}

Finally, if post-processing correction is applied to $\widehat{\bm{x}}$, the particles in the PF also need to be corrected accordingly: 
\begin{equation}
\label{equ:updatepf}
    \bm{x}^{(k)}_i = \widehat{\bm{x}}^{(k)}_i + \bm{x}^{(k)} - \widehat{\bm{x}}.
\end{equation}
If no post processing is performed, the output of the algorithm is simply 
$\bm{x}^{(k)} = \widehat{\bm{x}}$.

To summarize the \emph{Position Correction} part in PPC algorithm, we divide the position correction into three cases. (i) In Case 1, we find the optimal correction rotation angle and implement yaw correction; (ii) In Case 2, we treat this case as being blocked by an obstacle and implement step length correction; (iii) In Case 3, we assume the user has passed through the \emph{Door} and perform a \emph{Room} switch. The position correction process of PPC algorithm is illustrated in Fig.~\ref{fig:ppc}.

\begin{table*}[!t]
\footnotesize
\centering
\caption{Key Experimental Parameters}
\label{tab:ep}
\setlength{\extrarowheight}{1.5pt}
\setlength{\tabcolsep}{2.5mm}{
\begin{tabular}{|cc|c|cc|c|}
\hline
\multicolumn{2}{|c|}{\textbf{Device Parameter}} & \textbf{Value} & \multicolumn{2}{c|}{\textbf{FP-BP Algorithm Parameter}} & \textbf{Value} \\\hline\hline
\multicolumn{1}{|c|}{\multirow{4}{*}{BLE beacon}} & Model & RF-star RF-B-SR1 & \multicolumn{1}{c|}{\multirow{2}{*}{Floor plan}} & Resolution & $r=1/0.07~\mathrm{m^{-1}}$ \\\cline{2-3}\cline{5-6} 
\multicolumn{1}{|c|}{} & Broadcast interval &$200\,\mathrm{ms}$ & \multicolumn{1}{c|}{} & Virtual grid interval & $I_M = 0.3\m$ \\\cline{2-6} 
\multicolumn{1}{|c|}{} & Transmission power & $0\,\mathrm{dBm}$ & \multicolumn{1}{c|}{\multirow{3}{*}{GML}} & Estimation interval & $250~\mathrm{ms}$ \\\cline{2-3}\cline{5-6}
\multicolumn{1}{|c|}{} & Advertising mode & iBeacon & \multicolumn{1}{c|}{} & Kalman filter & $q_K=0.16$, $r_K=16$ \\\cline{1-3}\cline{5-6} 
\multicolumn{1}{|c|}{\multirow{6}{*}{Smartphone}} & Model & HONOR Magic 6 & \multicolumn{1}{c|}{} & Tunable parameters & $\kappa=0.01$, $\tau=0.5$ \\\cline{2-6}
\multicolumn{1}{|c|}{} & Mobile processor & Snapdragon 8 Gen3 & \multicolumn{1}{c|}{\multirow{3}{*}{PF fusion}} & Particle number & $m=500$ \\\cline{2-3}\cline{5-6}
\multicolumn{1}{|c|}{} & Operating system & Android 15 & \multicolumn{1}{c|}{} & Particle noise & $|\nu_s|\leqslant0.05\m$, $|\nu_\theta|\leqslant10^\circ$ \\\cline{2-3}\cline{5-6}
\multicolumn{1}{|c|}{} & RAM & 12 GB & \multicolumn{1}{c|}{} & Fusion variance & $\sigma_x^2=0.1$ \\\cline{2-6}
\multicolumn{1}{|c|}{} & BLE scan mode & Low latency & \multicolumn{1}{c|}{\multirow{2}{*}{PPC}} & Candidate angles & $\mathbf{\Phi}=\{\pm i\cdot5^\circ\}$, $1\leqslant i\leqslant12$ \\\cline{2-3}\cline{5-6}
\multicolumn{1}{|c|}{} & IMU sample rate & $60\,\mathrm{Hz}$ & \multicolumn{1}{c|}{} & Test scale factor & $f=1.5$ \\\hline
\end{tabular}
}
\end{table*}

The PPC algorithm is summarized in \textbf{Algorithm \ref{alg:PPC}}. Besides, it is worth noting that the number of consecutive occurrences of Case 2 should be counted. When Case 2 occurs multiple times consecutively, it indicates that the user has been blocked by an obstacle repeatedly. Since such a situation is unlikely to occur during normal walking, it is considered that the positioning result has deviated significantly and is therefore unreliable. In this case, the estimated result $\bm{x}^{(k)}$ should be corrected to $\bm{x}_B$ to avoid excessive errors.

\subsection{{Extending FP-BP for Multi-Floor Scenarios}}
{The above algorithm is designed for single-floor scenarios. In multi-floor positioning, users often move across floors via elevators or stairwells. We note that the proposed FP-BP can be readily extended to such scenarios.} 

{First, we can mark areas involving floor transitions (FTAs) in the floor plan, e.g., elevators and stairwells, with a new distinguishable color during the preprocessing step. This makes it straightforward to extract their binary mask $\bm{I}_{\textit{FTA}}$. Next, we can extend the binary mask $\bm{I}(\bm{x}^\mathrm{I})$ in \eqref{equ:mask} by defining $\bm{I} = \bm{I}_{\textit{Wall}} + 0.5\bm{I}_{\textit{Door}}+0.25\bm{I}_{\textit{FTA}}$, and the map $\bm{M}(\bm{x})$ in \eqref{equ:map} by defining $\bm{M}(\bm{x})=0.25$ if $\bm{x}$ is in the FTA. When the estimated position $\bm{x}^{(k)}$ enters the FTA, a floor determination is performed at each subsequent update, primarily based on BLE signals. Owing to their significant cross-floor attenuation\cite{Caso2020}, proximity to multiple beacons on another floor indicates a floor transition. In addition, if the smartphone is equipped with a barometer, the altitude can be estimated to further assist in floor determination\cite{Cock2021}.}

{Once the user is detected to have entered another floor, the system switches to the corresponding floor plan and re-initializes the algorithm, including resetting the initial position based on BLE and re-initializing the PF, so that a new round of single-floor positioning can be performed.}

\begin{figure}[!t]
\begin{subfigure}{0.33\textwidth}
    \includegraphics[width=\textwidth]{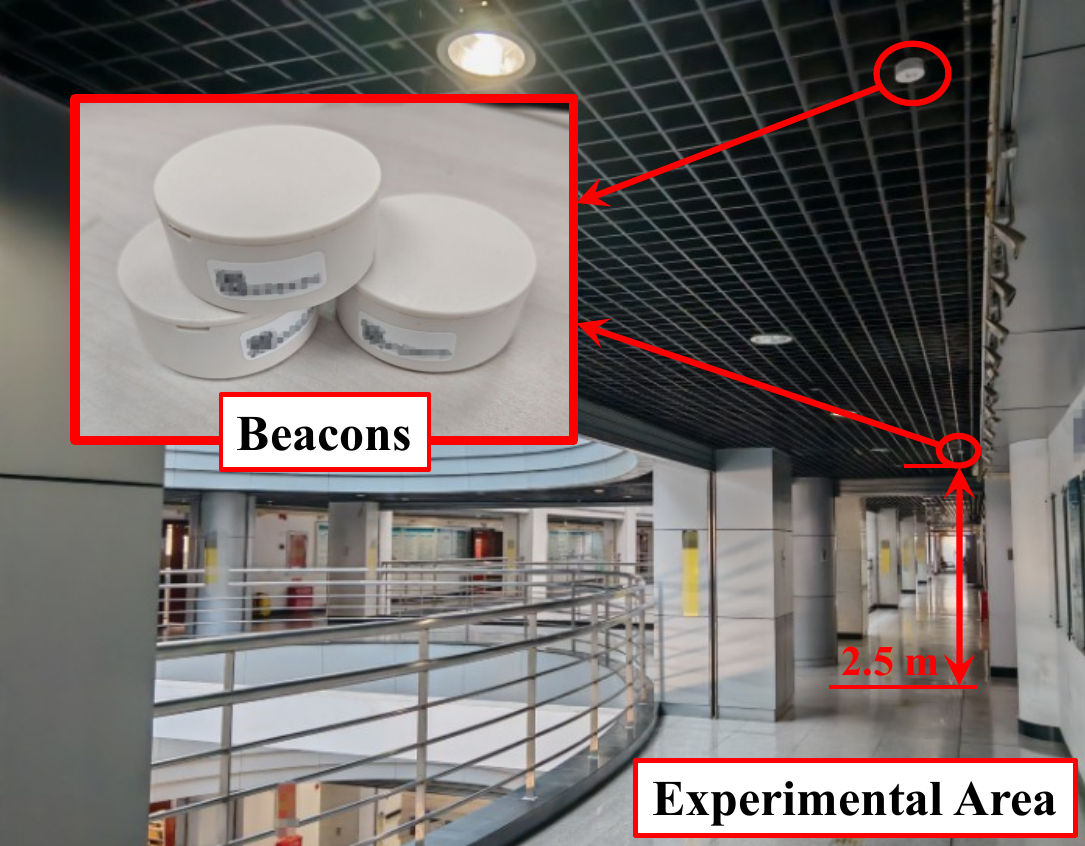}
    \centering
    \caption{}
    \label{fig:beacon_ex}
\end{subfigure}
\hspace{1mm}
\begin{subfigure}{0.122\textwidth}
    \centering
    \includegraphics[width=\textwidth]{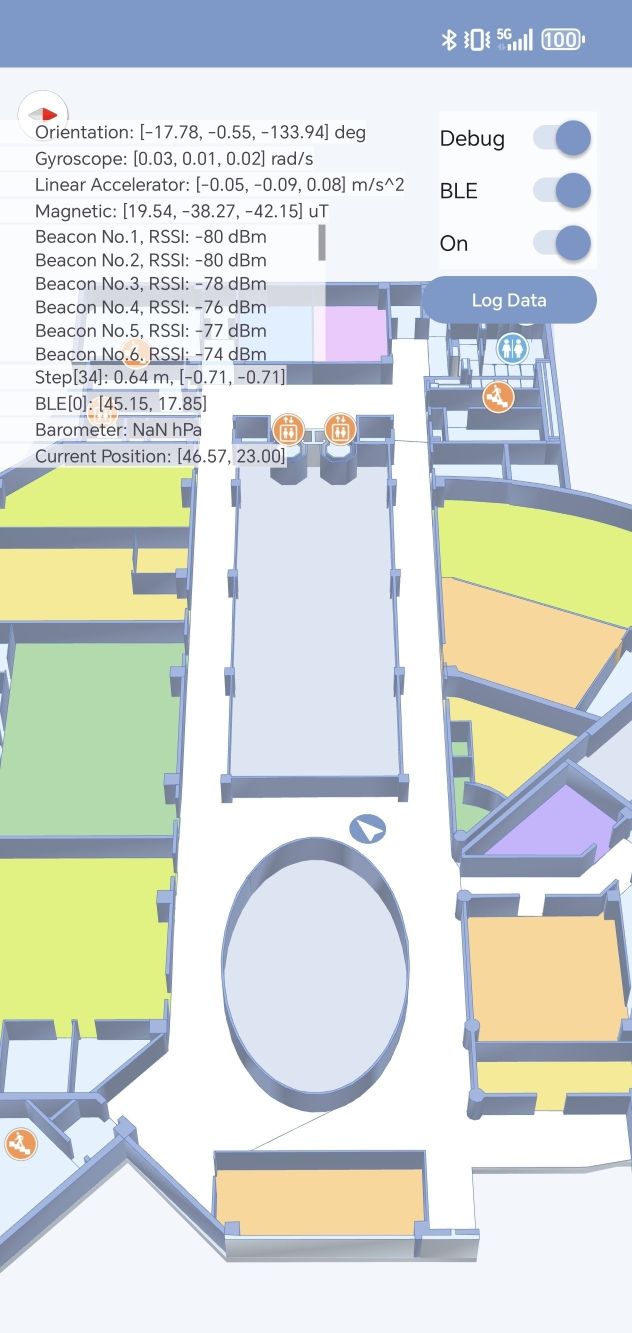}
    \caption{}
    \label{fig:phone}
\end{subfigure}
\centering
\caption{(a) BLE beacons in the experimental area; and (b) FP-BP prototype on the smartphone.}
\label{fig:proto}
\end{figure}

\begin{figure}[!t]
    \centering
    \includegraphics[width=0.475\textwidth]{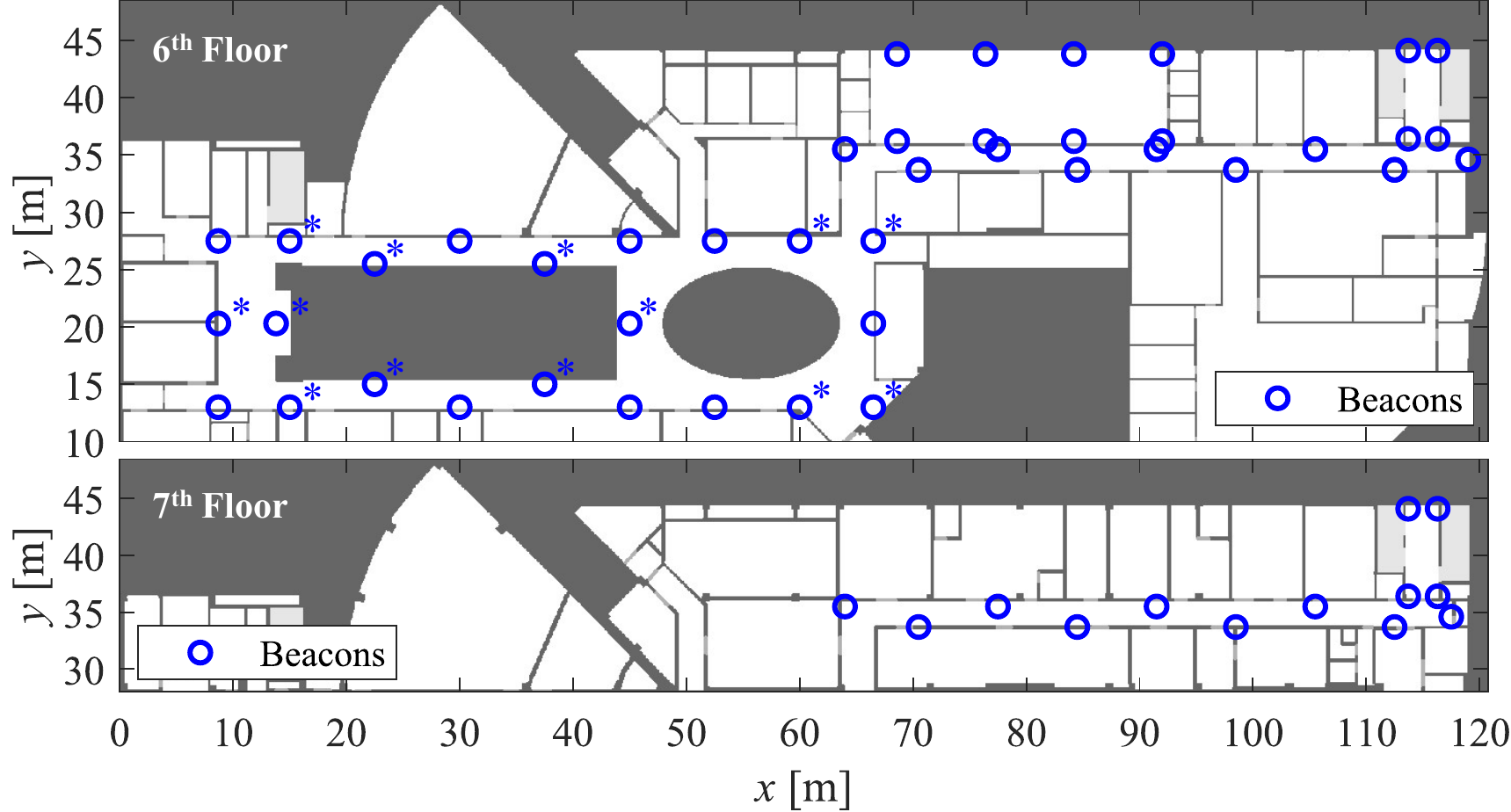}
    \caption{{BLE beacons in the experimental area. Only the beacons with ``\textcolor{blue}{\textbf{*}}" are considered in the experiments under the sparse beacon deployment.}}
    \label{fig:bs}
\end{figure}

\section{Experimental Setup and Results}
\label{sec:exp}

In this section, a prototype is established to implement FP-BP, and experiments are conducted to verify its performance and feasibility. Fig.~\ref{fig:proto} shows the experimental area and prototype of FP-BP, and key experimental parameters are listed in Table~\ref{tab:ep} unless otherwise specified. Next, we will detail the implementation of the experiments and analyze the results.

\subsection{Experimental Setup}

{The experimental area was located on the 6th and 7th floors of a teaching building, with a ceiling height of approximately $2.5\m$ and a total area of about $5760\m^2$ per floor. As shown in Fig.~\ref{fig:proto}\subref{fig:beacon_ex}, in our experiments, a total of $56$ beacons are deployed on the ceiling, which have distinct UUIDs and are deployed at the coordinates shown in Fig.~\ref{fig:bs}.}

\begin{figure}[!t]
\begin{subfigure}{0.475\textwidth}
    \includegraphics[width=\textwidth]{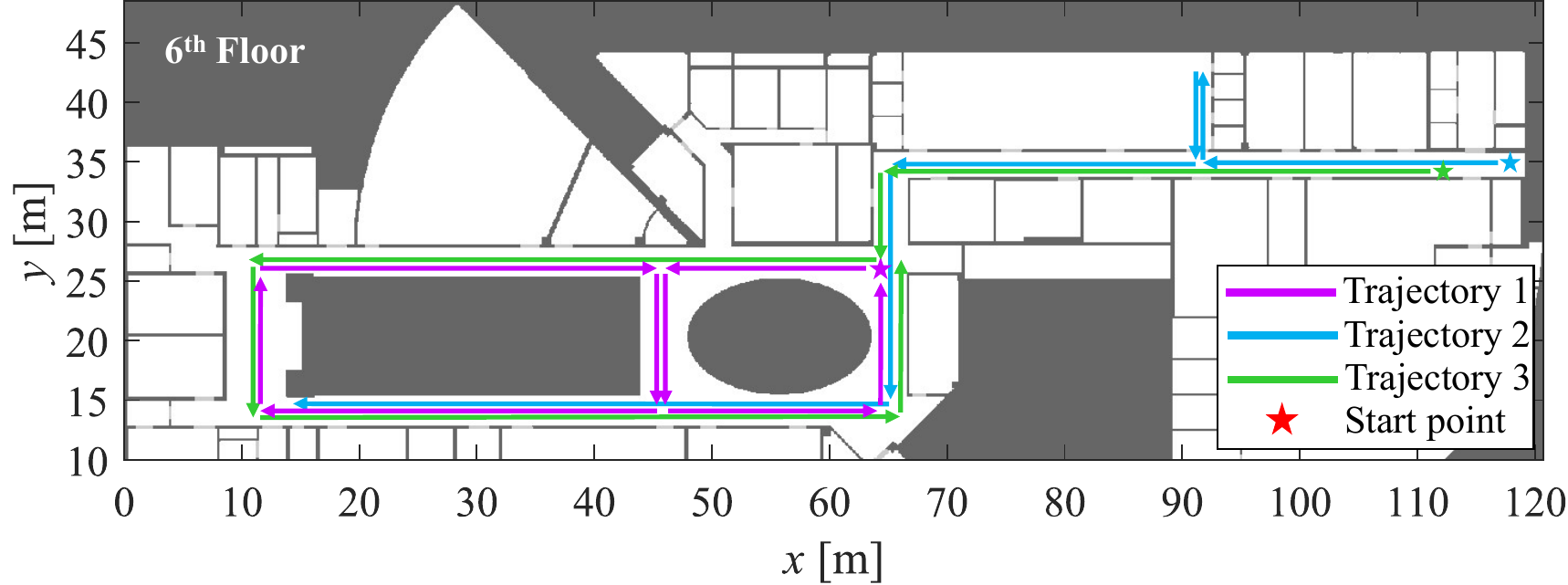}
    \centering
    \caption{{Trajectories 1--3.}}
    \label{fig:trajs_13}
\end{subfigure}
\centering
\begin{subfigure}{0.475\textwidth}
    \centering
    \includegraphics[width=\textwidth]{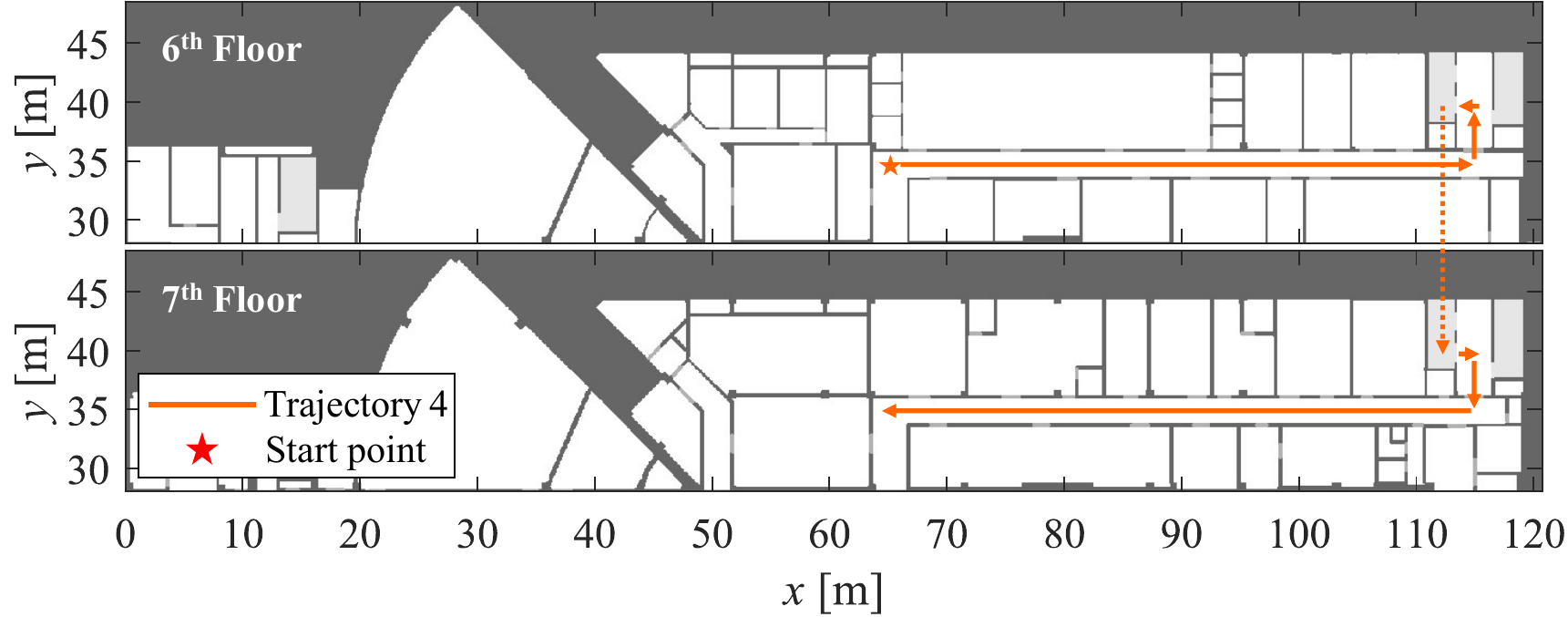}
    \caption{{Cross-floor trajectory 4.}}
    \label{fig:trajs_4}
\end{subfigure}
\centering
\caption{{Brief diagrams of experimental trajectories.}}
\label{fig:trajs}
\end{figure}

In the experiments, we set the tester to walk along a specific trajectory at a speed of approximately $1.0~\mathrm{m/s}$, holding the smartphone facing forward (i.e., the $y^\mathrm{D}$-axis of the phone is directed forward), with the step length fixed at $0.6\m$. {To verify the robustness of FP-BP, we set up four trajectories, as shown in Fig.~\ref{fig:trajs}. The total lengths of these trajectories are $316.8\m$, $137.4\m$, $187.8\m$ and $115.8\m$, respectively. In particular, for trajectory 1, the tester walked two laps for one trial; for trajectory 2, there are two \emph{Room} switch events, whereas other trajectories have none; and for trajectory 4, the tester started on the 6th floor and was transported by elevator to the 7th floor. The following Sections~\ref{ssec:bler} and \ref{ssec:fpbpr} present the results of $10$ trials of trajectory 1 conducted by two testers; Section~\ref{ssec:pmtr} presents the results and comparisons across these trajectories; and finally, Section~\ref{ssec:rcsm} compares the performance of FP-BP on different mobile devices.}

To evaluate the positioning performance, we first define the 2-D position error as the distance between the estimated position and the ground truth in MCS; then use the following metrics: (i) The mean of errors (MPE); (ii) 50\%/80\% cumulative distribution function (CDF) corresponding errors (P50/P80 scores); (iii) the standard deviation (STD) of errors; and (iv) the mean of multiple computation times (MCT).

\begin{figure*}[!t]
\begin{subfigure}{0.36\textwidth}
    \centering
    \includegraphics[height=0.59\textwidth]{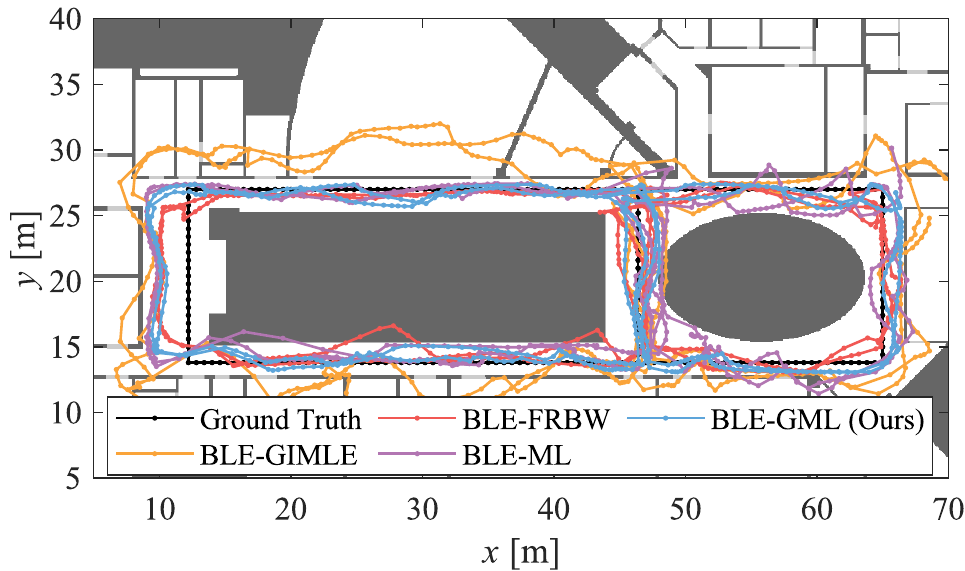}
    \caption{}
    \label{fig:ble_traj}
\end{subfigure}
\centering
\begin{subfigure}{0.265\textwidth}
    \centering
    \includegraphics[height=0.8\textwidth]{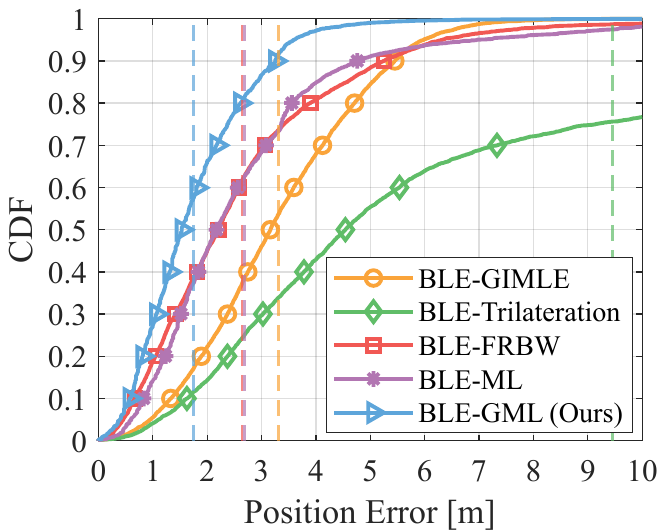}
    \caption{}
    \label{fig:ble_cdf}
\end{subfigure}
\centering
\begin{subfigure}{0.36\textwidth}
    \centering
    \includegraphics[height=0.59\textwidth]{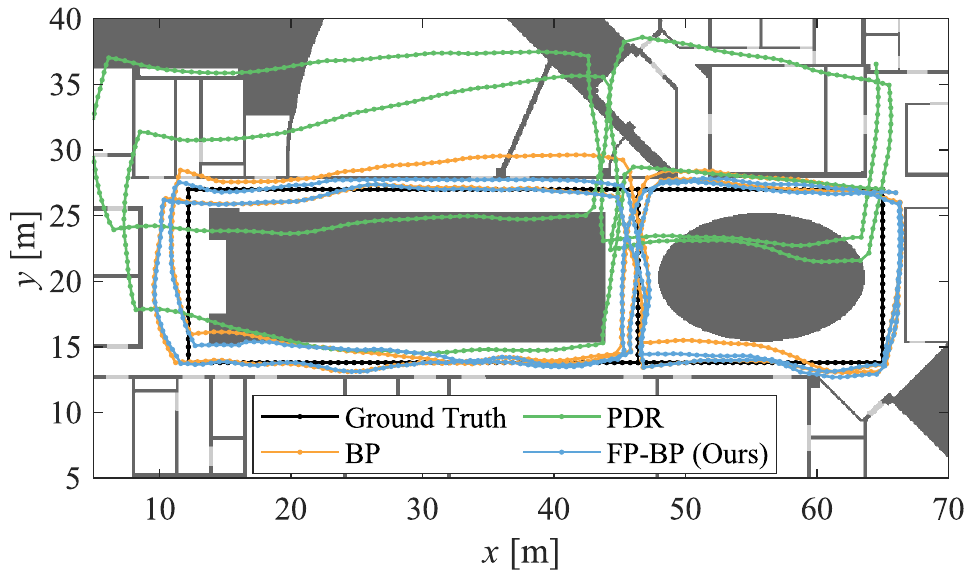}
    \caption{}
    \label{fig:my_traj}
\end{subfigure}
\centering
\begin{subfigure}{0.265\textwidth}
    \centering
    \includegraphics[height=0.8\textwidth]{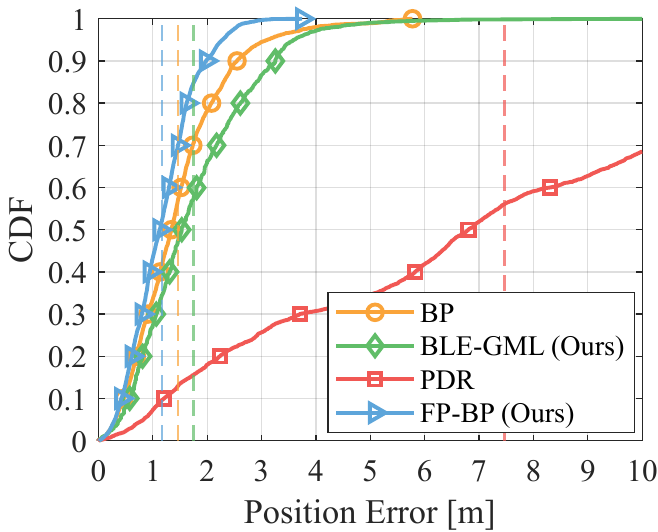}
    \caption{}
    \label{fig:my_cdf}
\end{subfigure}
\centering
\centering
    \begin{subfigure}{0.36\textwidth}
    \centering
    \includegraphics[height=0.59\textwidth]{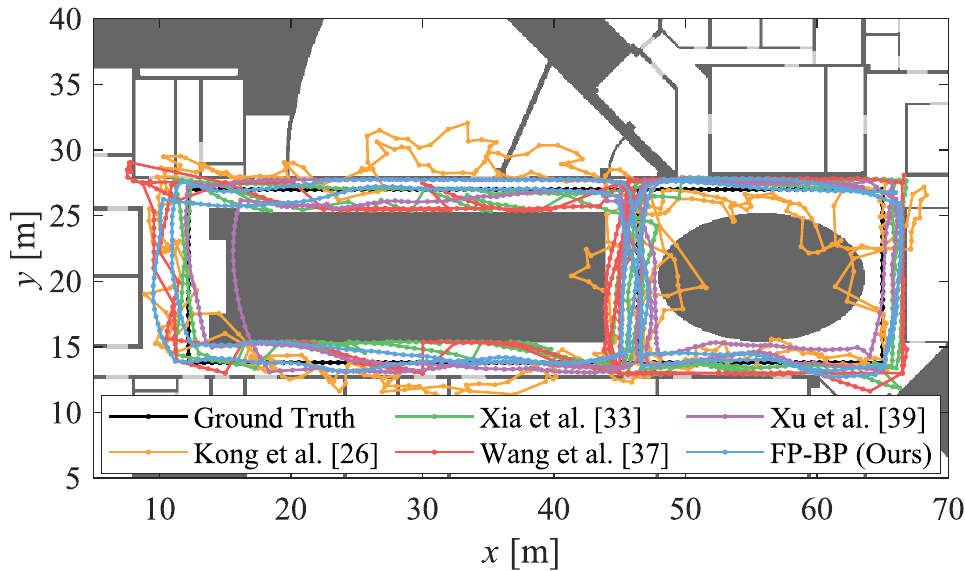}
    \caption{}
    \label{fig:bl_traj}
\end{subfigure}
\centering
\begin{subfigure}{0.265\textwidth}
    \centering
    \includegraphics[height=0.8\textwidth]{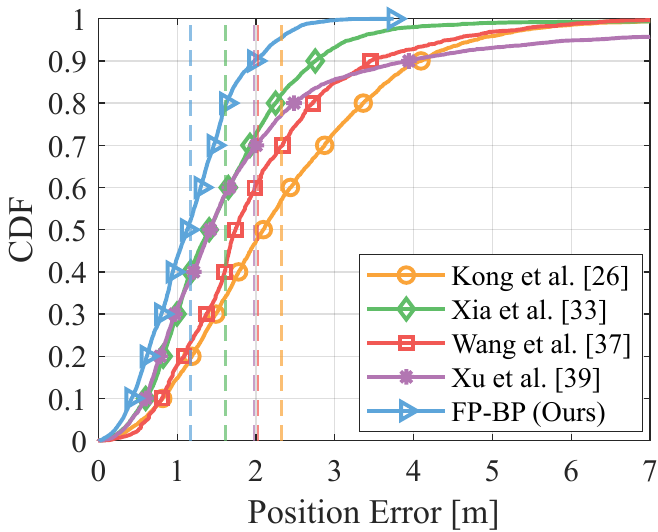}
    \caption{}
    \label{fig:bl_cdf}
\end{subfigure}
\centering
\caption{{Experimental results of trajectory 1. (a) and (b): Comparison of GML with BLE baseline algorithms. (c) and (d): Ablation comparison of FP-BP in different cases. (e) and (f): Comparison of FP-BP with baseline algorithms.}}
\centering
\label{fig:exp}
\vspace{-1em}
\end{figure*}

\subsection{BLE Positioning Results}
\label{ssec:bler}

\begin{figure}[t]
\centering
\includegraphics[width=0.275\textwidth]{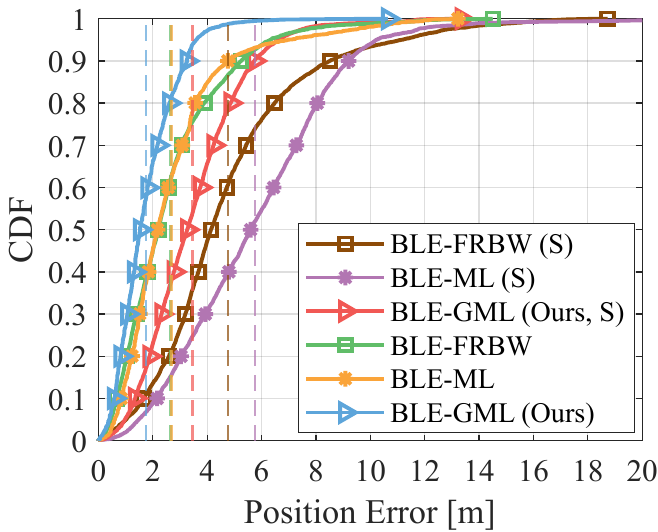}
\caption{{Experimental results of BLE positioning under normal (dense) and sparse beacon deployments. ``(S)" represents the sparse deployment.}}
\label{fig:exp_ble_s}
\end{figure}

We first analyze the experimental results of the proposed GML algorithm. {In terms of baselines, in addition to the traditional maximum likelihood (ML) algorithm in our previous work\cite{MyConf}, we also conduct another three common algorithms: a grid-based improved MLE (GIMLE) algorithm\cite{Feng2014}, the linear least squares (LLS)-based trilateration\cite{Poulose2019}, and a filtered RSSI and beacon weight (FRBW) algorithm\cite{Alsmadi2021}.} Note that: (i) The LLS-based trilateration takes all beacons into calculation. (ii) For the FRBW algorithm, the weight adjustable degree $g$ is set to $5$, and only the top $3$ beacons with the largest RSSIs are involved in the calculation. (iii) We apply the same mean filtering used in GML algorithm to the results of all algorithms. Taking the step moments detected by FP-BP as the reference, we record the BLE estimated position at each step moment, and calculate its error relative to the ground truth position.

The results are shown in Figs.~\ref{fig:exp}\subref{fig:ble_traj} and \subref{fig:ble_cdf}. Note that, since the accuracy of trilateration is significantly lower than that of the other methods, its trajectory is not presented in Fig.~\ref{fig:exp}\subref{fig:ble_traj}. {We can observe that despite using KF and mean filtering, the positioning results estimated by GIMLE still exhibit significant fluctuations, with some results deviating considerably from the ground truth. This is mainly because GIMLE does not take into account the geometric constraints formed by beacon positions.} In contrast, both GML, ML and FRBW algorithms successfully constrained their results within the experimental area. However, FRBW algorithm still has some estimated positions appearing inside obstacles, which is avoided by the proposed GML algorithm.

{Fig.~\ref{fig:exp}\subref{fig:ble_cdf} and Table~\ref{tab:exp_all} present the comparison of performance metrics between GML and baselines. We can see that GML algorithm achieves an MPE of $1.74\m$, which is lower than all the baselines, with a significant error reduction of over $34\%$. In addition, GML significantly outperforms all the baselines in terms of P50/P80 scores, and STD. These results demonstrate that the proposed GML algorithm achieves higher positioning accuracy and better stability.}

{In addition to the normal (dense) beacon deployment, we also consider a sparse deployment scenario in Fig.~\ref{fig:bs}. The performance under both deployments is compared in Fig.~\ref{fig:exp_ble_s}, using the CDFs of position error. Since ML and FRBW achieve the best performance in Table~\ref{tab:exp_all}, we select them as baselines. As shown, all algorithms experience degraded accuracy under the sparse deployment; however, GML still achieves the highest precision, with an MPE of $3.46\m$, outperforming ML ($5.76\m$) and FRBW ($4.77\m$). This demonstrates that GML maintains robust even in sparse layouts, since the constraint $\bm{y}\in\mathrm{int}\mathcal{C}_N$ in \eqref{equ:g1}, which is unsuitable for sparse deployments, is removed to avoid large deviations.}

\begin{table}[!t]
\footnotesize
\centering
\caption{{Experimental Results}}
\label{tab:exp_all}
\setlength{\extrarowheight}{1pt}
\setlength{\tabcolsep}{1.75mm}{
\begin{tabular}{|c|c|c|c|c|} \hline
\textbf{Methods} & \textbf{MPE [m]} & \textbf{P50 [m]} & \textbf{P80 [m]} & \textbf{STD [m]} \\\hline\hline
\textbf{GML (Ours)} & \textbf{1.74} & \textbf{1.53} & \textbf{2.61} & \textbf{1.08} \\\hline
GIMLE\cite{Feng2014} & 3.30 & 3.16 & 4.71 & 1.57 \\\hline
Trilateration\cite{Poulose2019} & 9.45 & 4.54 & 11.89 & 15.48 \\\hline
FRBW\cite{Alsmadi2021} & 2.64 & 2.21 & 3.90 & 1.99 \\\hline
ML\cite{MyConf} & 2.68 & 2.17 & 3.55 & 2.05 \\\hline\hline
\textbf{FP-BP (Ours)} & \textbf{1.14} & \textbf{1.09} & \textbf{1.62} & \textbf{0.59} \\\hline
BP & 1.46 & 1.34 & 2.08 & 0.91 \\\hline
PDR & 7.47 & 6.81 & 12.07 & 5.02 \\\hline
Kong et al.\cite{Kong2023} & 2.32 & 2.08 & 3.36 & 1.36 \\\hline
Xia et al.\cite{Xia2019} & 1.61 & 1.41 & 2.25 & 1.11 \\\hline
Wang et al.\cite{Wang2016An} & 2.02 & 1.74 & 2.72 & 1.22 \\\hline
Xu et al.\cite{Xu2019} & 1.97 & 1.42 & 2.48 & 1.91 \\\hline
\end{tabular}
}
\end{table}

\subsection{FP-BP Positioning Results}
\label{ssec:fpbpr}

To assess the performance of FP-BP, we conduct the fusion positioning systems proposed by Kong et al.\cite{Kong2023}, Xia et al.\cite{Xia2019}, Wang et al.\cite{Wang2016An} and Xu et al.\cite{Xu2019} as baselines for comparison. In particular: (i) The system proposed by Kong et al.\cite{Kong2023} is a tightly-coupled BLE/PDR fusion system without floor plan integration. (ii) Xia et al.\cite{Xia2019} introduce the floor plan into the traditional PF-based BLE/PDR fusion system. In their system, the weights of particles within obstacle areas are reset to zero. (iii) Wang et al.\cite{Wang2016An} identify landmarks such as access points, doors, corners, and walls in the floor plan, and implement specify position calibrations in their system. Note that due to the complexity of this floor plan, it is challenging to define the direction of corners. Therefore, only the positions of the corners are matched in our experiments. (iv) Xu et al.\cite{Xu2019} proposed an optimization scheme for particle re-initialization based on\cite{Xia2019}, taking into account the problem of particle depletion and result divergence. Note that we utilize the KF-processed BLE RSSIs in place of the WiFi fine timing measurement (FTM) to ensure a consistent comparison.

\begin{figure}[!t]
\centering
\includegraphics[width=0.42\textwidth]{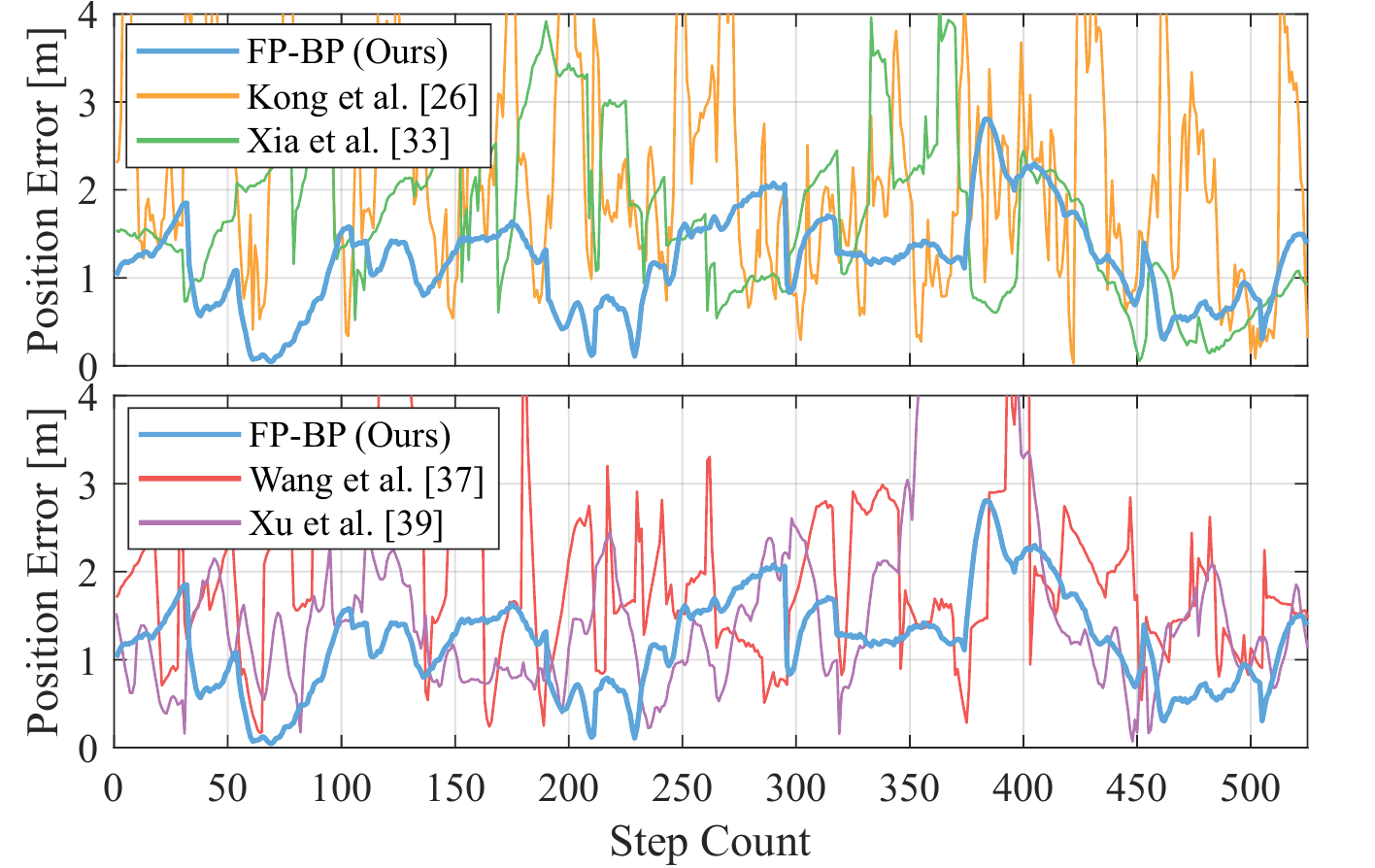}
\caption{PE versus step count for FP-BP and baseline algorithms.}
\label{fig:exp_steps}
\end{figure}

\begin{figure}[!t]
\centering
\includegraphics[width=0.385\textwidth]{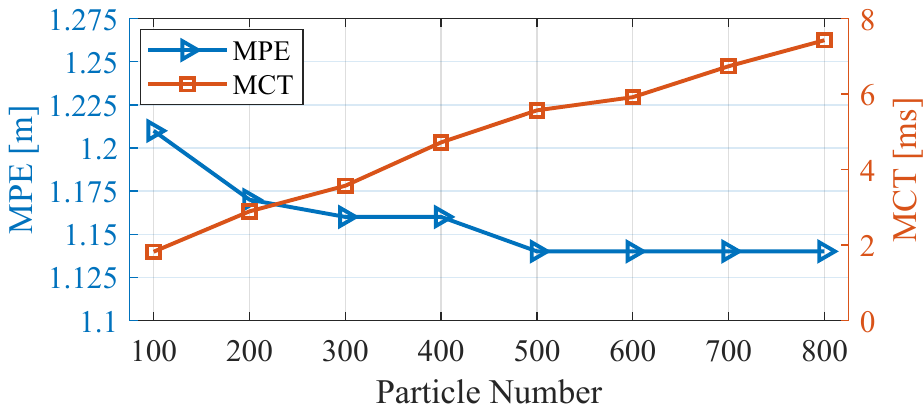}
\caption{{Comparison of MPE and MCT for different particle numbers.}}
\label{fig:exp_pf}
\end{figure}

The results are shown in Figs.~\ref{fig:exp}\subref{fig:my_traj}--\subref{fig:bl_cdf}, \ref{fig:exp_steps}, \ref{fig:exp_pf} and Table~\ref{tab:exp_all}. Figs.~\ref{fig:exp}\subref{fig:my_traj} and \subref{fig:my_cdf} are the ablation experiments which compare the results of FP-BP, FP-BP without PPC (denoted as BP), and PDR. It is evident that although the initial position of PDR is set as the ground truth, the error of PDR gradually increases as sensor errors accumulate, eventually deviating significantly from the ground truth. After fusing the BLE estimated position using PF, the cumulative error of BP is significantly reduced; however, many positioning results still ``pass through walls” and appear outside the intended walkable area. Finally, by applying the PPC algorithm to incorporate the floor plan, the FP-BP algorithm successfully confines these positioning points within the correct area.

{In Fig.~\ref{fig:exp}\subref{fig:my_cdf} and Table~\ref{tab:exp_all}, we can observe that the original PDR yields the largest MPE of $7.47\m$. After fusing BLE results, the MPE of BP is significantly reduced to $1.46\m$. With the further introduction of the proposed PPC for post-processing, the MPE of FP-BP is further reduced to $1.14\m$, representing an improvement of about $22\%$. In addition, BP fused with PDR also achieves higher accuracy than the pure BLE-based GML algorithm. In terms of P50/P80 scores and STD, FP-BP demonstrates the best positioning accuracy and stability. These results validate the effectiveness of each component in the proposed FP-BP fusion system.}

To further evaluate the performance of FP-BP, in Figs.~\ref{fig:exp}\subref{fig:bl_traj}, \subref{fig:bl_cdf} and \ref{fig:exp_steps}, we take the baseline systems\cite{Kong2023,Wang2016An,Xia2019} and\cite{Xu2019} for comparison. In Fig.~\ref{fig:exp}\subref{fig:bl_traj}, we can see that due to the lack of floor plan integration, some estimated positions of\cite{Kong2023} appear inside obstacles. Additionally, despite the incorporation of the floor plan, baseline\cite{Xu2019} still cannot prevent some results from erroneously appearing within obstacles. Other algorithms, including FP-BP, have successfully avoided this ``wall-crossing" problem.

From Fig.~\ref{fig:exp}\subref{fig:bl_cdf} and Table~\ref{tab:exp_all}, we can observe that among these algorithms, FP-BP achieves the lowest MPE at $1.14\m$, consistent with Fig.~\ref{fig:exp}\subref{fig:my_cdf}. In contrast, the MPEs of other baselines are all over $1.5\m$, which are significantly larger that of FP-BP, and FP-BP improves positioning accuracy by over $29\%$ compared with these baselines. Moreover, compared with the other baselines, FP-BP improves P50, P80, and STD by at least $23\%$, $28\%$, and $47\%$, respectively, demonstrating superior positioning performance.

\begin{figure}[!t]
\begin{subfigure}{0.475\textwidth}
    \includegraphics[width=\textwidth]{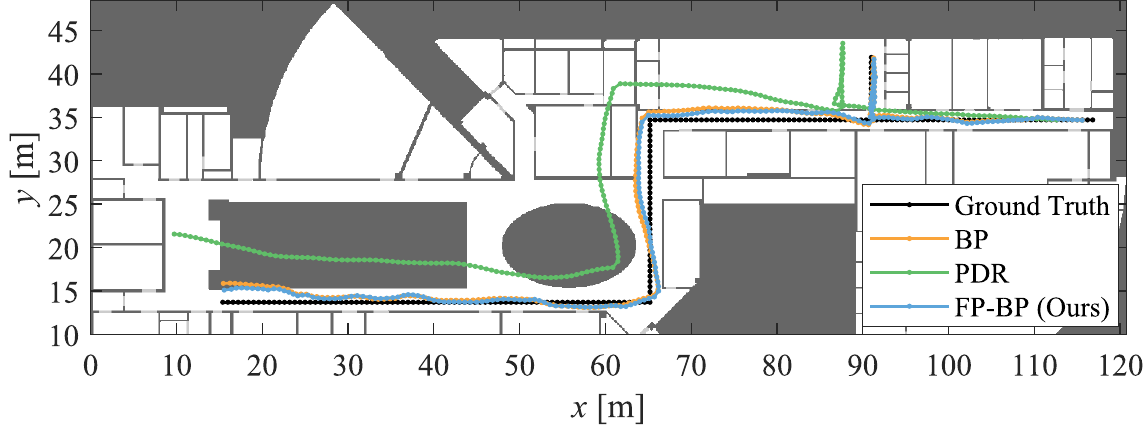}
    \centering
    \caption{Results of trajectory 2.}
    \label{fig:exp_traj2}
\end{subfigure}
\centering
\begin{subfigure}{0.475\textwidth}
    \centering
    \includegraphics[width=\textwidth]{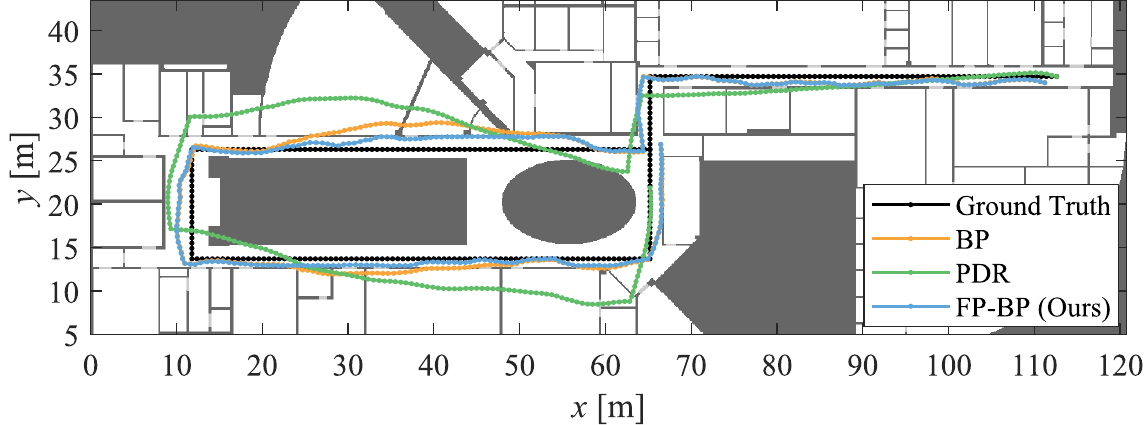}
    \caption{Results of trajectory 3.}
    \label{fig:exp_traj3}
\end{subfigure}
\centering
\begin{subfigure}{0.475\textwidth}
    \centering
    \includegraphics[width=\textwidth]{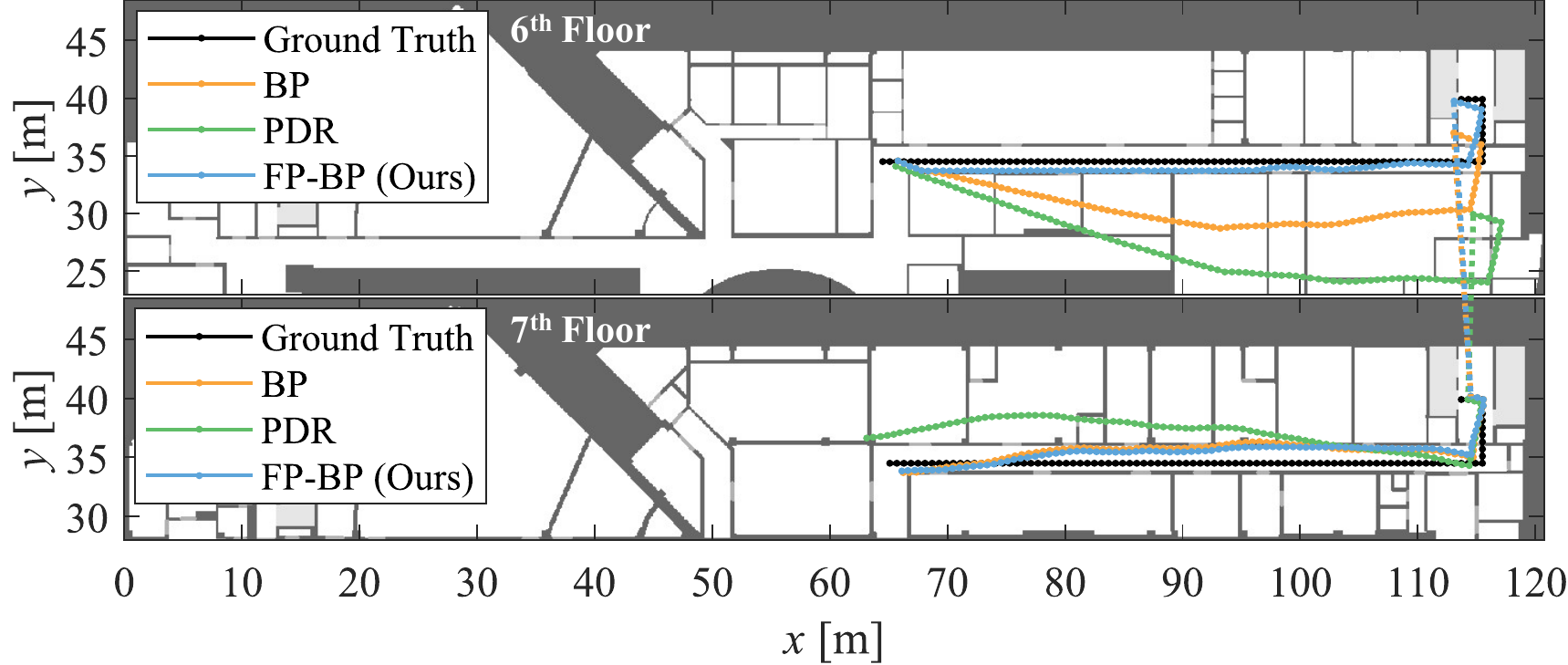}
    \caption{{Results of trajectory 4.}}
    \label{fig:exp_traj4}
\end{subfigure}
\centering
\caption{{Experimental results of FP-BP across trajectories.}}
\label{fig:exp_trajs}
\end{figure}

Fig.~\ref{fig:exp_steps} compares the errors for FP-BP and baseline algorithms as the step count increases in a single experiment. We can observe that all these algorithms effectively reduce cumulative errors, as their errors do not significantly increase with the step count. However, the results of\cite{Kong2023,Xia2019,Wang2016An} and\cite{Xu2019} exhibit significant fluctuations, with peak errors exceeding $3\m$, indicating that they struggle to provide stable positioning in certain areas. In contrast, the errors of FP-BP remain more stable, with about $90\%$ of errors kept within $2\m$. These results indicate that FP-BP demonstrates superior robustness compared to other fusion-based systems.

{In Fig.~\ref{fig:exp_pf}, we compare the MPE and MCT of FP-BP under different particle numbers. We can observe that as the particle number increases, the MPE decreases, which is consistent with \eqref{equ:mc}, since more particles can enable a more accurate estimation of posterior PDF $p(\bm{x}|\mathcal{Z})$. Meanwhile, the MCT increases with the particle number. When the particle number $m>500$, the MPE converges to $1.14\m$; therefore, we set $m=500$ and the corresponding MCT $<6\,\mathrm{ms}$. Furthermore, the proposed FP-BP requires significantly fewer particles than methods such as\cite{Choi2022} and\cite{Xu2019}, since the PPC mechanism only corrects the particle positions rather than discarding them, thus avoiding particle loss.}

\subsection{Results Comparison Across Trajectories}\label{ssec:pmtr}
\begin{figure}[!t]
\centering
\includegraphics[width=0.42\textwidth]{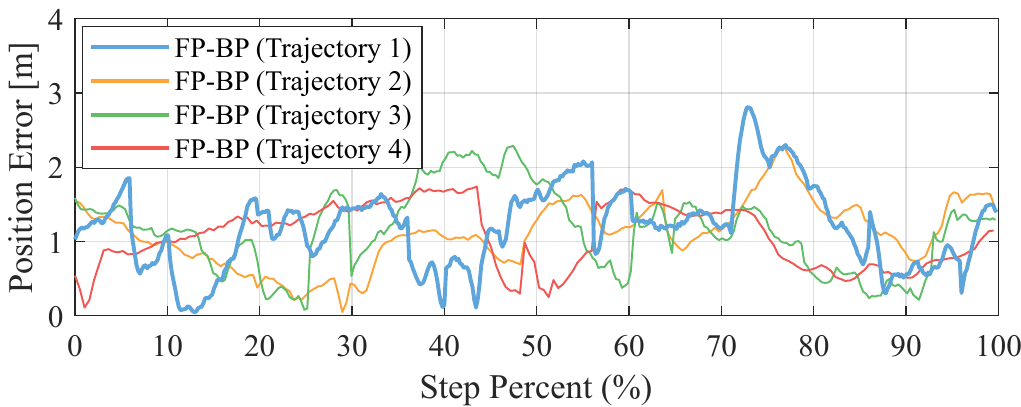}
\caption{{PE versus step percentage for FP-BP across trajectories.}}
\label{fig:exp_steps_traj}
\end{figure}
\begin{table}[!t]
\footnotesize
\centering
\caption{{Experimental Results Across Trajectories}}
\label{tab:exp_traj}
\setlength{\extrarowheight}{1.5pt}
\setlength{\tabcolsep}{1.3mm}{
\begin{tabular}{|ccc|cc|cc|cc|}
\hline
\multicolumn{3}{|c|}{\textbf{Trajectory}} & \multicolumn{2}{c|}{\textbf{MPE [m]}} & \multicolumn{2}{c|}{\textbf{P50 [m]}} & \multicolumn{2}{c|}{\textbf{P80 [m]}} \\\hline
\multicolumn{1}{|c|}{\textbf{No.}} & \multicolumn{1}{c|}{\textbf{Steps}} & \textbf{Dist. [m]} & \multicolumn{1}{c|}{\textbf{FP-BP}} & \textbf{BP} & \multicolumn{1}{c|}{\textbf{FP-BP}} & \textbf{BP} & \multicolumn{1}{c|}{\textbf{FP-BP}} & \textbf{BP} \\\hline\hline
\multicolumn{1}{|c|}{1} & \multicolumn{1}{c|}{528} & 316.8 & \multicolumn{1}{c|}{\textbf{1.14}} & 1.46 & \multicolumn{1}{c|}{\textbf{1.09}} & 1.34 & \multicolumn{1}{c|}{\textbf{1.62}} & 2.08 \\\hline
\multicolumn{1}{|c|}{2} & \multicolumn{1}{c|}{229} & 137.4 & \multicolumn{1}{c|}{\textbf{0.93}} & 0.98 & \multicolumn{1}{c|}{\textbf{0.92}} & 0.95 & \multicolumn{1}{c|}{\textbf{1.26}} & 1.38 \\\hline
\multicolumn{1}{|c|}{3} & \multicolumn{1}{c|}{313} & 187.8 & \multicolumn{1}{c|}{\textbf{0.80}} & 1.13 & \multicolumn{1}{c|}{\textbf{0.74}} & 0.98 & \multicolumn{1}{c|}{\textbf{1.23}} & 1.49 \\\hline
\multicolumn{1}{|c|}{4} & \multicolumn{1}{c|}{193} & 115.8 & \multicolumn{1}{c|}{\textbf{1.07}} & 2.13 & \multicolumn{1}{c|}{\textbf{1.12}} & 1.65 & \multicolumn{1}{c|}{\textbf{1.49}} & 3.62 \\\hline
\end{tabular}
}
\end{table}

To verify the robustness of FP-BP under various trajectories, we further conduct comparative experiments across another three trajectories, as shown in Figs.~\ref{fig:exp_trajs}, \ref{fig:exp_steps_traj} and Table~\ref{tab:exp_traj}. In Figs.~\ref{fig:exp_trajs}\subref{fig:exp_traj2} and \subref{fig:exp_traj3}, we can see that similar to Fig.~\ref{fig:exp}\subref{fig:my_cdf}, FP-BP effectively constrains the results within the correct \textit{Room}. Moreover, when the user passes through a \textit{Door} into another \textit{Room}, it accurately handles this \textit{Room} switch event. {Fig.~\ref{fig:exp_trajs}\subref{fig:exp_traj4} shows the results of cross-floor trajectory 4. We can see that the extended FP-BP can effectively handle multi-floor scenarios, and no additional error is introduced during floor transitions. This is because the algorithm is re-initialized upon reaching the new floor, thereby eliminating the accumulated error from the previous floor.}

In Fig.~\ref{fig:exp_steps_traj}, we can observe that across all trajectories, as the step count increases, FP-BP similarly effectively mitigates cumulative errors, keeping almost all the errors within $2\m$. Additionally, {Table~\ref{tab:exp_traj} indicates that FP-BP demonstrates excellent positioning performance across trajectories 2--4, achieving MPEs of $0.93\m$, $0.80\m$ and $1.07\m$ respectively. In contrast, the BP system without floor plan integration reaches MPEs of $0.98\m$, $1.13\m$ and $2.13\m$. Furthermore, the P50 and P80 scores of FP-BP are also less than those of BP.} These results indicate that FP-BP operates stably in indoor environments, remaining unaffected by changes in trajectory, and exhibits both high accuracy and strong robustness.

\subsection{{Results Comparison Across Smartphones}}\label{ssec:rcsm}

\begin{table}[!t]
\footnotesize
\centering
\caption{{Smartphone Parameters}}
\label{tab:smt}
\setlength{\extrarowheight}{1.5pt}
\setlength{\tabcolsep}{0.8mm}{
\begin{tabular}{|c|c|c|c|c|}
\hline
\textbf{No.} & \textbf{Model} & \textbf{Mobile processor} & \textbf{Operating system} & \textbf{RAM}\\\hline\hline
1 & {\scriptsize HONOR Magic 6} & {\scriptsize Snapdragon 8 Gen3} & Android 15 & 12 GB \\\hline
2 & {\scriptsize HUAWEI Mate 40 Pro} & {\scriptsize Kirin 9000} & Android 10 & 8 GB \\\hline
3 & {\scriptsize realme Q2i} & {\scriptsize Dimensity 720} & Android 10 & 4 GB \\\hline
\end{tabular}
}
\end{table}
\begin{figure}[!t]
\centering
\includegraphics[width=0.38\textwidth]{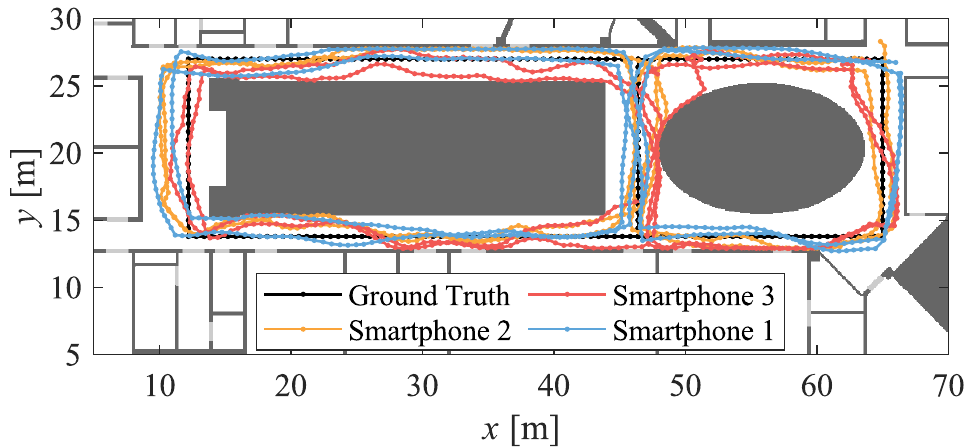}
\caption{{Experimental results of FP-BP across smartphones.}}
\label{fig:exp_smt}
\end{figure}
\begin{table}[!t]
\footnotesize
\centering
\caption{{Experimental Results Across Smartphones}}
\label{tab:exp_smt}
\setlength{\extrarowheight}{1.5pt}
\setlength{\tabcolsep}{0.7mm}{
\begin{tabular}{|c|c|c|c|c|c|c|}
\hline
\textbf{No.} & \textbf{MPE [m]} & \textbf{P50 [m]} & \textbf{P80 [m]} & \textbf{STD [m]} & \textbf{MCT\textsubscript{1} [ms]} & \textbf{MCT\textsubscript{2} [ms]}\\\hline\hline
1 & 1.14 & 1.09 & 1.62 & 0.59 & 5.57 & 4.67 \\\hline
2 & 1.57 & 1.36 & 2.38 & 0.95 & 2.76 & 6.31 \\\hline
3 & 1.63 & 1.50 & 2.27 & 0.82 & 5.94 & 7.67 \\\hline
\end{tabular}
}
\end{table}

{The performance of PDR-based algorithms is typically closely related to the quality of smartphone sensors. To verify the feasibility of FP-BP across different smartphones, in addition to smartphone 1 in Table~\ref{tab:ep}, two additional smartphones shown in Table~\ref{tab:smt} were introduced. These smartphones are equipped with different mobile processors and random access memories (RAMs). We conducted experiments along trajectory 1 using each of these smartphones, and the results are presented in Fig.~\ref{fig:exp_smt} and Table~\ref{tab:exp_smt}. The trajectories in Fig.~\ref{fig:exp_smt} indicate that the proposed FP-BP algorithm yields consistent results across all three smartphones without significant deviation. The statistical results in Table~\ref{tab:exp_smt} show that FP-BP performs best on smartphone 1. Although its performance slightly decreases on smartphones 2 and 3, the MPE remains around $1.5\m$, and the STD is below $1\m$, suggesting that the accuracy is stable and consistent across different positions.}

{In addition, in Table~\ref{tab:exp_smt}, let MCT\textsubscript{1} and MCT\textsubscript{2} denote the MCT of \textit{GML} module and \textit{PDR\&Fusion} module, respectively. We can observe that all MCTs are at the millisecond level, which are far below the BLE signal broadcasting and position estimation intervals. This demonstrates that the proposed FP-BP algorithm imposes a low computational burden on the device, offering strong feasibility for achieving low-latency real-time positioning.}

\section{Conclusion}
\label{sec:con}
In this paper, we have proposed the FP-BP algorithm, which can fully utilize the floor plan information to achieve real-time fusion positioning without limitations on map geometry. We have also proposed the GML algorithm to enhance RSSI-based BLE positioning. In particular, based on the accurate extraction of map features, the floor plan is preprocessed in the offline phase to construct an indoor map. In the online phase, based on the MLE principle, we apply the GML algorithm to obtain BLE results. Then, smartphone sensor data is fused by a PF to perform an initial estimation. Finally, based on the preprocessed indoor map, floor plan information is integrated by applying the proposed PPC mechanism to the initial estimation. {To evaluate the performance of FP-BP, we conducted both single- and cross-floor experiments in a building with a total area of over 5,000 m\textsuperscript{2} per floor. The results indicate that the FP-BP algorithm can achieve a mean positioning accuracy of 1.14 m, outperforming other floor plan-fused baselines by over 29\%.} Moreover, FP-BP algorithm has also demonstrated superior real-time performance on mobile devices. Therefore, it holds great promise for large-scale indoor positioning. {In future work, we aim to extend FP-BP to support diverse user carrying poses and motion patterns, further improving its adaptability to real-world scenarios.}


\bibliographystyle{IEEEtran}
\bibliography{references}

 




\vfill

\end{document}